\documentclass[aps,prl,twocolumn,groupedaddress,superscriptaddress,showpacs]{revtex4}

\usepackage{amsmath, amsthm, amssymb}

\usepackage{graphicx}
\usepackage{psfrag}
\usepackage{dsfont}
\usepackage{tabularx}
\usepackage{hhline}

\newcolumntype{x}[1]{>{\centering\arraybackslash\hspace{0pt}}p{#1}}

\usepackage{xcolor}
\definecolor{mydarkgreen}{rgb}{0.0,0.5,0.0}

\usepackage[linktocpage=true,
  colorlinks=true, 
  pdfborder={0 0 0},
  linkcolor=blue,
  citecolor=red,
  filecolor=yellow,
  urlcolor=mydarkgreen,
  bookmarks,
  pdftitle={FT-RDMFT: Correlation},
  pdfauthor={Tim Baldsiefen},
 pdfkeywords={Reduced density matrix functional theory,electron gas},
]{hyperref} 

\usepackage{breakurl}

\newcommand{\tb}[1]{\textbf{#1}}
\newcommand{\tr}[1]{\textup{tr}\{#1\}}
\newcommand{\eq}[1]{(\ref{#1})}

\newcommand{\ket}[1]{{|#1\rangle}}
\newcommand{\bra}[1]{{\langle#1|}}

\newcommand{\nn}{\nonumber}
\newcommand{\veps}{\varepsilon}

\newcommand{\abref}[1]{#1}

\begin{document}


\title{Reduced density matrix functional theory at finite temperature. III. \\Application to the electron gas: Correlation effects}


\author{Tim Baldsiefen}
\affiliation{Institut f\"ur Theoretische Physik, Freie Universit\"at Berlin, Arnimallee 14, D-14195 Berlin, Germany}
\affiliation{Max-Planck-Institut f\"ur Mikrostrukturphysik, Weinberg 2, D-06112 Halle, Germany}
\author{E. K. U. Gross}
\affiliation{Max-Planck-Institut f\"ur Mikrostrukturphysik, Weinberg 2, D-06112 Halle, Germany}


\date{\today}

\begin{abstract}
Based on our derivation of finite temperature reduced density matrix functional theory \cite{Baldsiefen_al_1.2012} and the discussion of the performance of its first-order functional \cite{Baldsiefen_al_2.2012} this work presents several different correlation-energy functionals and applies them to the homogeneous electron gas. The zero temperature limits of the correlation-energy and the momentum distributions are investigated and the magnetic phase diagrams in collinear spin configuration are discussed.
\end{abstract}

\pacs{31.15.ec,31.15.E-,65.40.-b,71.10.Ca}

\maketitle


\section{Introduction}

In Part I of the present work \cite{Baldsiefen_al_1.2012} we have presented the theoretical foundations of finite temperature reduced density matrix functional theory (FT-RDMFT). FT-RDMFT and its zero temperature counterpart employ the one-reduced density matrix (1RDM) rather than the density as central variable and are therefore also suitable for the description of systems subject to nonlocal external potentials. Furthermore, the availability of the eigenvalues and eigenstates of the 1RDM, which are many-particle objects in contrast to the Kohn-Sham energies and states, opens up the possibility to describe phenomena that are difficult to access by DFT \cite{Helbig_al.2007,Marques_Lathiotakis.2008,Piris_al.2010,Helbig_Lathiotakis_Gross.2009,Baldsiefen.2010,Lathiothakis_al.2009,Sharma_al.2008}. One prominent example are Mott insulators \cite{Sharma_al.2008} which also establish the need for a finite temperature version of RDMFT.

In view of the extraordinary success of the LDA in the context of DFT, it seems desirable to pursue a similar approach in the context of FT-RDMFT. If one considers only local external potentials then there exists a big variety of methods to calculate the equilibrium properties of the homogeneous electron gas (HEG) at either zero \cite{Singwi_Tosi_Land_Sjolander.1968,Ceperley_Alder.1980,Perdew_Wang.1992,Dharma-wardana_Perrot.2000} or finite temperatures \cite{Tanaka_Ichimaru.1986,Schweng_Boehm.1993,Iyetomi_Ichimaru.1986,Tanaka_Ichimaru.1989,Perrot_Dharma-Wardana.2000,Zong_Lin_Ceperley.2002,Conduit_Green_Simons.2009,Perrot.1979,Gupta_Rajagopal_2.1980,Dharma-Wardana_Taylor.1981,Gupta_Rajagopal.1982,Perrot_Dharma-Wardana.1984,Kanhere_Panat_Rajagopal_Callaway.1986,Dandrea_Ashcroft_Carlsson.1986}. However, for an LDA in FT-RDMFT one needs the equilibrium free energy of the HEG subject to arbitrary nonlocal external potentials. These could in principle be calculated by Monte-Carlo calculations which, to our knowledge, have not been carried out so far. Accordingly, for setting up an LDA in FT-RDMFT, one needs to work with functionals that are approximate even for the HEG. Part II of this series \cite{Baldsiefen_al_2.2012} has therefore investigated the first-order functional of FT-RDMFT when applied to the HEG and it was shown that it was capable of reproducing a qualitatively acceptable phase diagram including collinear as well as spiral spin configurations. The resulting critical densities, however, where far too large which emphasizes the need for approximate correlation-energy functionals in FT-RDMFT.

The main goal of this third part of our work is the investigation of correlation functionals in FT-RDMFT when applied to the HEG. This will be done as follows. Firstly we will review the most relevant concepts of FT-RDMFT, discussing the possiblity of the formulation of a functional analoguous to the LDA. Secondly we will use Monte-Carlo results to investigate the relation between correlation induced by electrons of the same spin compared to correlation resulting from the interaction of electrons with different spins. These results will rule out the most common class of approximations in unpolarized RDMFT as candidates for an accurate description of spin polarized systems and they will be used to guide the development of improved functionals. Thirdly, we will resort to the zero temperature situation and propose a functional which for the first time not only reproduces the correlation-energy of the HEG over a wide range of densities but also describes the momentum distribution qualitatively correctly. Finally, we will use the perturbative methodology as presented in \cite{Baldsiefen_al_1.2012} to derive a truly temperature dependent correlation functional in FT-RDMFT.

\section{Theoretical foundations}
We will now briefly review the most important concepts of FT-RDMFT \cite{Baldsiefen_al_1.2012}. 
A general quantum mechanical ensemble is described by its statistical density operator (SDO) $\hat D$, which is a weighted sum of projection operators on the underlying Hilbert space:
\begin{align}
  \hat D&=\sum_iw_i\ket{\Psi_i}\bra{\Psi_i},\quad w_i\geq0,\sum_iw_i=1\label{eq.sdo}.
\end{align}
The corresponding 1-reduced density matrix (1RDM) is defined as the spatial contraction of all but one dimensions of $\hat D$
\begin{align}
  \gamma(\tb r,\tb r')&=\tr{\hat D\hat\Psi^+(\tb r')\hat\Psi(\tb r)},
\end{align}
where the density $\rho$ is determined via $\rho(\tb r)=\gamma(\tb r,\tb r)$. As the 1RDM, by construction, is hermitian we can write it in spectral representation as
\begin{align}
  \gamma(\tb r,\tb r')&=\sum_in_i\phi_i^*(\tb r')\phi_i(\tb r),
\end{align}
where the eigenvalues $\{n_i\}$ are usually called occupation numbers (ONs) and the eigenstates $\{\phi_i\}$ are called natural orbitals (NOs) \cite{Loewdin.1955}.

We have shown in \cite{Baldsiefen_al_1.2012} that the grand potential of an arbitrary system with possibly nonlocal external potential in grand canonical equilibrium (eq) is uniquely determined by the corresponding eq-1RDM. The grand potential $\Omega$ of a system with 1RDM $\gamma$ can therefore be written as a functional with a universal part $\mathfrak{F}[\gamma]$, which is independent of the external potential of the system, and a part $\Omega_{ext}[\gamma]$ which just depends on this external influence.
\begin{align}
  \Omega[\gamma]&=\mathfrak{F}[\gamma]+\Omega_{ext}[\gamma].
\end{align}
The external functional $\Omega_{ext}[\gamma]$ is given by
\begin{align}
  \Omega_{ext}[\gamma]&=\int d\tb rd\tb r'(v_{ext}(\tb r',\tb r)-\mu\delta(\tb r-\tb r'))\gamma(\tb r,\tb r'),\label{eq.gp._ext}
\end{align}
where the chemical potential $\mu$ governs the coupling to the particle bath. The universal functional $\mathfrak{F}[\gamma]$ on the other hand is defined as
\begin{align}
  \mathfrak{F}[\gamma]&=\min_{\hat D\rightarrow\gamma}\left(\tr{\hat D(\hat T+\hat W-1/\beta\ln\hat D)}\right)\label{eq.f_univ}\\
  &=\Omega_k[\gamma]-1/\beta S_0[\gamma]+\Omega_H[\gamma]+\Omega_x[\gamma]+\Omega_c[\gamma]
\end{align}
where $\beta=1/(k_B T)$ describes the coupling to the energy bath with $k_B$ and $T$ being Boltzmann's constant and the temperature respectively. The different functionals for the kinetic contribution $\Omega_k[\gamma]$, the Hartree and exchange contributions $\Omega_H[\gamma],\Omega_x[\gamma]$ and the noninteracting entropy part $S_0[\gamma]$ are explicit functionals of the 1RDM and are given as
\begin{align}
  \Omega_{k}[\gamma]&=\int d\tb r\lim_{\tb r\rightarrow\tb r'}\left(-\frac{\nabla^2}{2}\right)\gamma(\tb r,\tb r')\\
  \Omega_{H}[\gamma]&=\frac12\int d\tb rd\tb r'w(\tb r',\tb r)\gamma(\tb r',\tb r')\gamma(\tb r,\tb r)\\
  \Omega_{x}[\gamma]&=-\frac12\int d\tb rd\tb r'w(\tb r',\tb r)\gamma(\tb r,\tb r')\gamma(\tb r',\tb r)\label{eq.rdmft.ex}\\
  S_0[\gamma]&=-\sum_i\left(n_i\ln(n_i)+(1-n_i)\ln(1-n_i)\right)\label{eq.rdmft.ent}
\end{align}
Accordingly, one only has to approximate the remaining correlation functional $\Omega_c[\gamma]$. As the nonlocal nature of the 1RDM allows an exact description of the kinetic energy the correlation functional only contains contributions from the interaction and the entropy which is expected to simplify the construction of approximate functionals.

The construction of the Kohn-Sham (KS) system in FT-RDMFT was shown to be particulary simple. The NO form the eigenstates of the KS Hamiltonian and the eigenenergies are found by an inversion of the Fermi-Dirac relation:
\begin{align}
  n_i&=\frac{1}{1+e^{\beta(\veps_i-\mu)}}\\
  \veps_i-\mu&=\frac{1}{\beta}\ln\left(\frac{1-n_i}{n_i}\right)\label{eq.ks.ee}.
\end{align}

With these findings we have all necessary tools at hand to describe the thermodynamic variables of a quantum system in grand canonical equilibrium. There are, however, physical problems which require the formulation in terms of a canonical ensemble, e.g. the question of the equilibrium magnetic moment of a solid. In this situation the number of electrons is fixed and the equilibrium is realized by the state of minimal free energy $F=E-1/\beta S$ rather than minimal grand potential.

As already mentioned in Part I of this work \cite{Baldsiefen_al_1.2012} the Hohenberg-Kohn theorem for canonical ensembles follows exactly the same steps as for the grand canonical ensemble with the only difference that the external potential is uniquely determined by the 1RDM only up to an additional constant. Just as in the grand canonical case the free energy functional $F[\gamma]$ can be seperated in a universal part $\mathfrak{F}^c[\gamma]$ and an external part $F_{ext}[\gamma]$ where the universal part is defined as in Eq. \eq{eq.f_univ} but the SDOs are now only defined on the $N$-particle Hilbert space. The external part is defined as in Eq. \eq{eq.gp._ext}, but without the chemical potential contribution. If we now assume our system to be in the thermodynamic limit, the functional for the free energy is simply related to the grand potential functional via
\begin{align}
  F[\gamma]&=\Omega[\gamma]+\mu N[\gamma].
\end{align}
This is due to the fact that in the thermodynamic limit the energy and entropy are identical for both canonical and grand canonical ensembles and therefore the universal functionals $\mathfrak{F}^c[\gamma]$ and $\mathfrak{F}[\gamma]$ coincide. The external functional describes a one-particle contribution and therefore only depends on the 1RDM, regardless of the ensemble under investigation.

After these considerations we can find the eq-free energy $F_{eq}$ of a certain phase by a minimization of $F[\gamma]$.
\begin{align}
  F_{eq}&=\min_{\gamma}F[\gamma]
\end{align}
The domain of minimization is given by the set of all ensemble-N-representable 1RDMS, i.e. 1RDMs whose ONs fulfill
\begin{align}
  0\leq n_i\leq1,\quad\sum_in_i=N,
\end{align}
with a fixed particle number $N$. Further details of the numerical treatment of FT-RDMFT for the HEG can be found in \cite{Baldsiefen_al_2.2012}.

We have shown in Part II of this work \cite{Baldsiefen_al_2.2012}, that the method of \abref{FT-RDMFT} constitutes an efficient tool for the description of the equilibrium-properties of the \abref{HEG} at finite temperature in \abref{FT-HF} approximation. Eventually, however, we would like to be able to describe correlation in spatially nonuniform systems, a task we will deal with in the following.

\section{Correlation in FT-RDMFT}
Motivated by the big success of the LDA and LSDA in DFT, the formulation of a local-1RDM-approximation (LRDMA) in FT-RDMFT seems desirable. At this point it becomes important to point out the conceptual differences between RDMFT and DFT.

\subsection{Local 1RDM approximation (LRDMA)}
In DFT, one only has to calculate the correlation-energy of the HEG as a function of the constant density $\rho$. In RDMFT, on the other hand, one has to calculate it for all possible density matrices $\gamma(\tb r-\tb r')$ that are compatible with translational invariance, i.e. for all possible momentum distributions $n(\tb k)$. The correlation-energy per volume of the HEG in RDMFT therefore becomes a \textit{functional}, $\veps_c^{HEG}[\gamma(\tb k,\tb R)]$, of $n(\tb k)$ rather than just a function of $\rho$. Under the assumption that such a functional was accessible, the LRDMA of RDMFT is defined in the following way.
\begin{align}
  E_c^{LRDMA}[\gamma(\tb r,\tb r')]&:=\int d^3 R\veps_c^{HEG}[\gamma(\tb k,\tb R)],\label{eq.lrdma}
\end{align}
where $E_c^{LRDMA}$ is the approximate correlation-energy of a nonuniform system and $\gamma(\tb k,\tb R)$ describes the Wigner transform of the 1RDM,
\begin{align}
  \gamma(\tb k,\tb R)&=\int d^3s\gamma(\tb R+\tb s/2,\tb R-\tb s/2)e^{i\tb k\cdot\tb s}.
\end{align}
This procedure is similar to the definition of the LDA in superconducting DFT \cite{Kurth_Marques_Lueders_Gross.1999,Ullrich_Gross.1996}. While there are many LDA constructions conceivable that correctly reduce to the homogeneous limit, Eq. \eq{eq.lrdma} is the only definition that correctly reproduces the correlation energy of a \textit{weakly inhomogeneous} electron gas \cite{Ullrich_Gross.1996}.

A parametrization of $\veps_c^{HEG}[n(\tb k)]$ has not been carried out so far and remains an important task for the future. An extension of the LRDMA fomalism to spin-dependent systems and systems at finite temperature is conceptually straightforward. However, the explicit calculation of the temperature-dependent $\veps_c^{HEG}[n(\tb k),T]$, although in principle possible by means of path integral Monte-Carlo techniques \cite{Ceperley.1995}, is rather involved, due to the fermionic sign problem \cite{Troyer_Wiese.2005}.

An alternative approach for further theoretical development in RDMFT and FT-RDMFT therefore consists in the derivation of approximate correlation functionals for the HEG and their implementation in a (FT-)LRDMA.
We point out that a functional in FT-RDMFT should not only reproduce an accurate eq-free energy but also has to yield a good momentum distribution.

The main focus of this work is therefore the presentation and investigation of several choices of correlation functionals in the theoretical framework of \abref{FT-RDMFT}. 

\subsection{Correlation-energy from DFT}

One choice which immediately suggests itself is the utilization of correlation functionals from DFT, i.e. functionals which just depend on the density $\rho$, rather than the full 1RDM:
\begin{align}
  \Omega_c[\gamma]&=\Omega_c[\rho].
\end{align}
We would like to point out that this approach exhibits an important intrinsic flaw. This becomes clear, if one reviews the minimization procedure for the case of local external potentials and seperates the variation over 1RDMs in a combined variation over densities and a variation over the class of 1RDMs which yield a certain density:
\begin{align}
  F_{eq}&=\min_{\gamma}F[\gamma]=\min_{\rho}\min_{\gamma\rightarrow \rho}F[\gamma]\\
  &=\min_{\rho}\Big(\Omega_{ext}[\rho]+\Omega_{H}[\rho]+\Omega_c[\rho]\nn\\
  &\hspace*{15mm}+\min_{\gamma\rightarrow \rho}\left(\Omega_k[\gamma]-k_BTS_0[\gamma]+\Omega_x[\gamma]\right)\Big).
\end{align}
As we assumed the correlation contribution to be independent of the particular form of the 1RDM as long as the diagonal, i.e. the density, stays the same we can take it out of the minimization over 1RDMs. We find that the minimization over the 1RDMs now only contains kinetic, exchange, and entropy contributions. If the density $\rho$ refers to a solution of the finite temperature Hartree-Fock equations to some external potential then we have shown in Part I  of this work \cite{Baldsiefen_al_1.2012} that the constrained minimization of this combination of functionals gives the corresponding Hartree-Fock momentum distribution. The resulting 1RDM  will in this sense be ``uncorrelated''. The severe problem with this result can be understood by considering the zero-temperature limit. Any choice of correlation functional in FT-RDMFT which just depends on the density will lead to a step function as the momentum distribution for low temperatures \footnote{This argument hinges on the assumption that the density $\rho$ is a Hartree-Fock density of some external potential. In the case of the HEG this is obviously justified but for general systems with nonlocal external potential it is not clear if all densities share this property. Nonetheless the previous considerations show that for a wide class of external potentials, i.e. all which lead to a groundstate density which is the Hartree-Fock density to some other external potential, the optimal 1RDM will be uncorrelated.}. This knowledge of the nature of the failure of DFT-functionals allows us to, at least partially, take account of these shortcomings and we will be able to study the effect of two representative DFT-functionals later on in our investigation of the magnetic phasediagram of the HEG in FT-RDMFT.

We have seen in \cite{Baldsiefen_al_2.2012} that the first-order functional in FT-RDMFT was capable of reproducing a qualitatively resonable phase diagram. Apart from the incorrect momentum distributions the main problem of these results was that the quantum phase transition between unpolarized and polarized configurations is instantaneous at zero temperature which is not correct for a quantum phase transition and that they occur at densities which are far too big ($r_s=5.56 a.u.$ compared to $r_s=50 a.u.\sim75a.u.$ \cite{Ceperley_Alder.1980,Zong_Lin_Ceperley.2002}). We concluded that the noninteracting entropy functional is capable of describing the most important finite temperature features of the magnetic phase diagram, including a weakening of the ferromagnetic phase and the reproduction of a critical temperature $T_c$, close to the Fermi temperature. As a first step in the development of correlation functionals in FT-RDMFT, it therefore seems desirable to construct a zero temperature correlation functional which yields a critical density close to the expected one and then investigate the temperature dependence induced by the noninteracting entropy functional. But before we are going to investigate such functionals we will first focus on an important property of the correlation-energy of a spin-polarized HEG.

\subsection{Spin-channel seperability}\label{sec.rdmft.magnetism}
The energy differences between different magnetic phases of the \abref{HEG} become very close for low densities. An error in the spin-dependence of the correlation-energy might therefore result in a big error in the estimate of the critical density. Therefore, even in Monte-Carlo calculations, the estimates for the critical Wigner-Seitz radius $r_c$, where a quantum phase transition between an unpolarized and totally polarized configuration occurs, vary considerably (e.g. $r_s=50a.u.$:\cite{Ceperley_Alder.1980} compared to $r_s=75a.u.$:\cite{Zong_Lin_Ceperley.2002}). An approximate functional in \abref{RDMFT} would need to exhibit a very high accuracy to be able to reproduce the quantum phase transition in the \abref{HEG}, which seems to be a very hard problem to solve. However, most physical systems of interest exhibit much higher densities and therefore the accurate description of the low density quantum phase transition of the HEG is rather irrelevant for the application of an LDA-type functional to real-world systems. 
This serves as an argument that for the implementation in an LDA also a medium accuracy in the description of the correlation-energy would suffice. In the following, we will elucidate, why the use of a certain, rather general class of functionals will have severe problems to achieve an even mediocre accuracy in describing spin-polarized systems.

When trying to describe spin-polarized systems in collinear configuration, the kinetic, the external potential, and the exchange part can be seperated into a sum of contributions from only spin-up and only spin-down \abref{NO}s and \abref{ON}s. To describe the \abref{xc} functional for a spin-polarized system, a common approach in \abref{RDMFT} is to make the same ansatz of spin-channel seperability.
\begin{align}
  E_{xc}[\gamma]&=E_{xc}[\gamma_{\uparrow\uparrow},\gamma_{\downarrow\downarrow}]=E^S_{xc}[\gamma_{\uparrow\uparrow}]+E^S_{xc}[\gamma_{\downarrow\downarrow}],
\end{align}
where $\gamma_{\uparrow\uparrow}$ and $\gamma_{\downarrow\downarrow}$ are the diagonal elements of the \abref{1RDM} $\gamma$ in spin-space and $E^S[\gamma]$ denotes the functional which just depends on one spin-component of the 1RDM. 

We claim that such an approximation is intrinsically incapable of describing both spin-polarized and spin-unpolarized configurations together. The underlying reason for this problem is that an additive functional might describe the correlation contributions, which arise from interactions of the electrons of the same spin (i.e. equal-spin-channel correlation), well but will not be able to describe the contribution coming from the correlation of electrons of different spin (i.e. opposite-spin-channel correlation). In the following, we will elaborate on this problem by considering a spin-polarized \abref{HEG}.  We will again rely on the PWCA\cite{Perdew_Wang.1992} parametrization of the polarization-dependent correlation energies.

For a collinear spin configuration, the fundamental quantities are the spin-up density $n_\uparrow$ and the spin-down density $n_\downarrow$ and the energy can be written as $E(n_\uparrow,n_\downarrow)$. If the assumption of spin-cannel seperability was valid, the following two relations would hold.
\begin{align}
  E(n_\uparrow,n_\downarrow)&=E(n_\uparrow,0)+E(0,n_\downarrow)=:E^P\label{eq.rdmft.sep.ef}\\
  E(n_\uparrow,n_\downarrow)&=\frac12\left(E(n_\uparrow,n_\uparrow)+E(n_\downarrow,n_\downarrow)\right)=:E^U\label{eq.rdmft.sep.ep}
\end{align}
In Eq. \eq{eq.rdmft.sep.ef} the partially polarized system is given as a sum of two fully polarized systems, whereas in Eq. \eq{eq.rdmft.sep.ep} one constructs the partially polarized one out of two unpolarized systems, hence the notations $E^P$ and $E^U$.

\begin{figure}[t!]
  \psfrag{tx}[][][1][0]{$r_s(a.u.)$}
  \psfrag{ty}[b1][b1][1][0]{$\Delta^P(Ha)$}
  \psfrag{tt}{\hspace*{-0mm}}
  \psfrag{t0}{\hspace*{-0mm}0}
  \psfrag{t2}{\hspace*{-0mm}0.2}
  \psfrag{t4}{\hspace*{-0mm}0.4}
  \psfrag{t6}{\hspace*{-0mm}0.6}
  \psfrag{t8}{\hspace*{-0mm}0.8}
  \psfrag{tz}{\hspace*{-0mm}1.0}
  \includegraphics[width=.8\columnwidth]{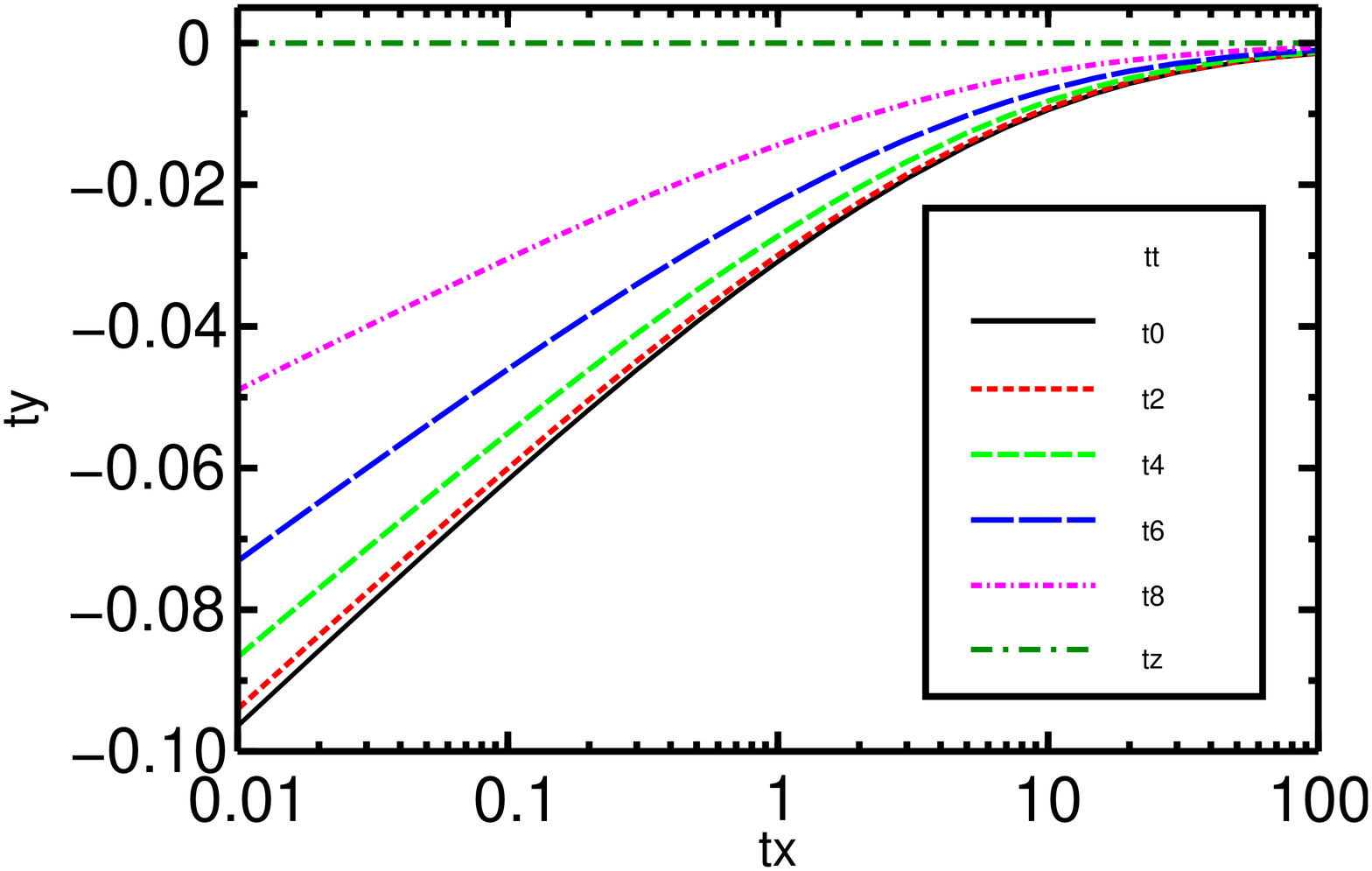} 
  \caption{Energy differences $\Delta^P$ from Eq. \eq{eq.rdmft.sep.diff-f} for different polarizations}\label{fig.rdmft.sep.deltaf}
\end{figure}

\begin{figure}[t!]
  \psfrag{tx}[][][1][0]{$r_s(a.u.)$}
  \psfrag{ty}[b1][b1][1][0]{$\Delta^U(Ha)$}
  \psfrag{tt}{\hspace*{-0mm}}
  \psfrag{t0}{\hspace*{-0mm}0}
  \psfrag{t2}{\hspace*{-0mm}0.2}
  \psfrag{t4}{\hspace*{-0mm}0.4}
  \psfrag{t6}{\hspace*{-0mm}0.6}
  \psfrag{t8}{\hspace*{-0mm}0.8}
  \psfrag{tz}{\hspace*{-0mm}1.0}
  \includegraphics[width=.8\columnwidth]{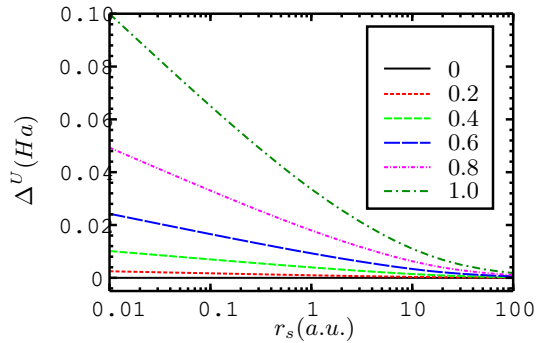} 
  \caption{Energy differences $\Delta^U$ from Eq. \eq{eq.rdmft.sep.diff-p} for different polarizations}\label{fig.rdmft.sep.deltap}
\end{figure}

We can now investigate, whether or not Eqs. \eq{eq.rdmft.sep.ef} and \eq{eq.rdmft.sep.ep} are valid for the case of a partially polarized \abref{HEG} by calculating the differences between the exact, i.e. Monte-Carlo, results and the expected results $E^P$ and $E^U$.
\begin{align}
  \Delta^P&=E-E^P\label{eq.rdmft.sep.diff-f}\\
  \Delta^U&=E-E^U\label{eq.rdmft.sep.diff-p}
\end{align}
The results are shown in Figures \ref{fig.rdmft.sep.deltaf} and \ref{fig.rdmft.sep.deltap}.

As we can see in Figure \ref{fig.rdmft.sep.deltaf}, $\Delta^P$ vanishes by construction for a totally polarized system ($\xi=1$). Decreasing the polarization then leads to a decrease of $\Delta^P$, i.e. $E<E^P$. As in $E^P$ the opposite-spin-channel correlation is neglected, we deduce that it has to be negative. Considering the unpolarized case in Figure \ref{fig.rdmft.sep.deltap}, we see that again by construction $\Delta^U$ vanishes, now for the unpolarized configuration ($\xi=0$) and then increases with increasing polarization. This can be understood by realizing that $E^U$ basically double counts the opposite-spin-channel correlation contribution, leading to $E>E^U$.

\begin{figure}[t!]
  \psfrag{tx}[][][1][0]{$r_s(a.u.)$}
  \psfrag{ty}[][][1][0]{$\delta^P$}
  \psfrag{tt}{\hspace*{-0mm}$\xi$}
  \psfrag{t0}{\hspace*{-0mm}0}
  \psfrag{t2}{\hspace*{-0mm}0.2}
  \psfrag{t4}{\hspace*{-0mm}0.4}
  \psfrag{t6}{\hspace*{-0mm}0.6}
  \psfrag{t8}{\hspace*{-0mm}0.8}
  \psfrag{tz}{\hspace*{-0mm}1.0}
  \includegraphics[width=.8\columnwidth]{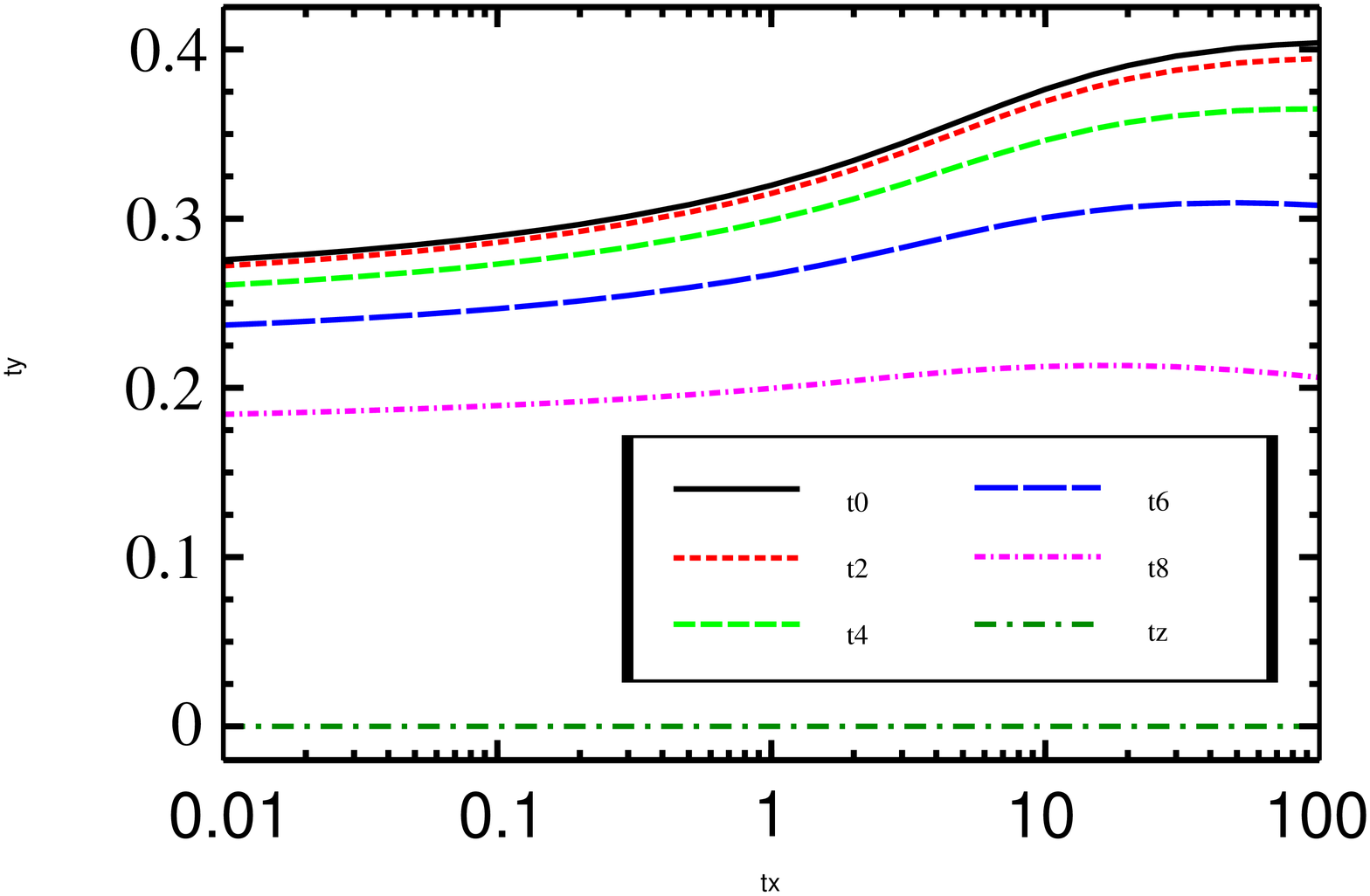} 
  \caption{Relative energy deviations $\delta^P$ from Eq. \eq{eq.rdmft.sep.frac-f} for different polarizations.}\label{fig.rdmft.sep.fracf}
\end{figure}

\begin{figure}[t!]
  \psfrag{tx}[][][1][0]{$r_s(a.u.)$}
  \psfrag{ty}[][][1][0]{$\delta^U$}
  \psfrag{tt}{\hspace*{-0mm}$\xi$}
  \psfrag{t0}{\hspace*{-0mm}0}
  \psfrag{t2}{\hspace*{-0mm}0.2}
  \psfrag{t4}{\hspace*{-0mm}0.4}
  \psfrag{t6}{\hspace*{-0mm}0.6}
  \psfrag{t8}{\hspace*{-0mm}0.8}
  \psfrag{tz}{\hspace*{-0mm}1.0}
  \includegraphics[width=.8\columnwidth]{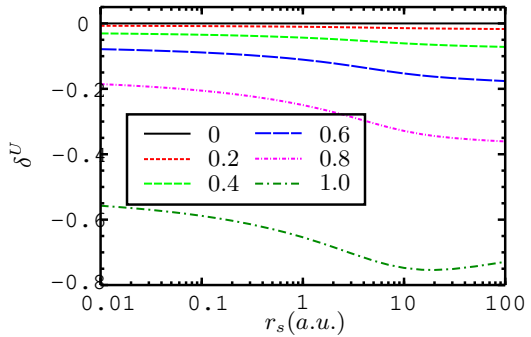} 
  \caption{Relative energy deviations $\delta^U$ from Eq. \eq{eq.rdmft.sep.frac-p} for different polarizations}\label{fig.rdmft.sep.fracp}
\end{figure}

In \abref{RDMFT}, the only correlation contribution comes from the interaction alone. We therefore would like to know, how the opposite-spin-channel correlation contribution relates to this correlation contribution $W_c$. We can use the \abref{PWCA} parametrization of the correlation contributions to the energy to calculate $W_c=E_c-T_c$ and investigate the following two fractions.
\begin{align}
  \delta^P&=\frac{\Delta^P}{W_c}\label{eq.rdmft.sep.frac-f}\\
  \delta^U&=\frac{\Delta^U}{W_c}\label{eq.rdmft.sep.frac-p}
\end{align}
The results are shown in Figures \ref{fig.rdmft.sep.fracf} and \ref{fig.rdmft.sep.fracp}. We see that by assuming spin-channel seperability, one will  yield errors for the correlation-energy of up to 40\%. However, a remarkable feature of Figures \ref{fig.rdmft.sep.fracf} and \ref{fig.rdmft.sep.fracp} is that the relative deviations over the whole range of considered densities do not vary strongly. Apparently, both equal- as well as opposite-spin-channel correlation are affected in the same way by a change in the density, leading only to a small change in their fraction. 

After these preconsiderations we are now prepared to investigate the performance of zero temperature RDMFT when applied to the HEG.

\subsection{Zero Temperature RDMFT}
Most energy functionals in \abref{RDMFT} try to incorporate correlation via a modification of the exchange functional Eq. \eq{eq.rdmft.ex}. The resulting functionals are accordingly called exchange-correlation functionals and can generally be written as
\begin{align}
  E_{xc}[\gamma]&=E_{x}[\gamma]+E_{c}[\gamma]\\
  E_{xc}[\gamma]&=-\frac12\sum_{ij}f(n_i,n_j)\nn\\
  &\hspace*{6mm}\int dxdx'w(x,x')\phi_i^*(x')\phi_i(x)\phi_j^*(x)\phi_j(x')\label{eq.rdmft.exc}.
\end{align}
Choosing $f^x(n_i,n_j)=n_in_j$ reproduces the exchange-only functional, neglecting correlation completely. 

The first approximation to $f(n_i,n_j)$ was done by M\"uller in 1984 \cite{Mueller.1984,Buijse_Baerends.2002} leading to $f^{Mueller}(n_i,n_j)=\sqrt{n_in_j}$. This M\"uller functional was able to correctly describe the dissociation limit of several small dimers of open-shell atoms but it overestimates the correlation-energy quite considerably. Inveigled by the simplicity of the form of the \abref{xc} approximation and the primary success of the M\"uller functional, Gritsenko, Pernal and Baerends developed what is now known as the BBC1, BBC2 and BBC3 functionals \cite{Gritsenko_Pernal_Baerends.2005}. The key difference in their approach is the different treatment of orbitals which were occupied or unoccupied in a Hartree-Fock solution (termed strongly and weakly occupied in the following). In addition, BBC2 and BBC3 effectively mix parts of the exchange and M\"uller functionals. A very similar approach was taken by Piris et al. \cite{Piris_al.2010}. Their approach differs slightly in the distinction of strongly and weakly occupied orbitals, removes parts of the self interaction (PNOF0) and tries to incorporate particle-hole symmetry (PNOF). These more elaborate functionals BBC1/2/3 and PNOF/0 are capable of reproducing good dissociation energies as well as correlation energies.

Whereas the previous functionals are mainly derived from physical arguments, Marques and Lathiotakis \cite{Marques_Lathiotakis.2008} pursued a different way by proposing a two-parameter Pad\'{e} form for $f(n_i,n_j)$. These parameters are then optimized to minimize the deviation of correlation energies when applied to the molecules of the G2 and G2-1 sets. The resulting functionals are called ML and ML-SIC, either including or excluding self interaction. The ML and ML-SIC functionals achieve an unprecedented precision, reaching the accuracy of second-order M\o ller-Plesset perturbation theory for the calculation of the correlation energies. An overview of several of these functionals as well as the corresponding energies for many molecules can be found in Ref. \cite{Lathiotakis_Marques.2008}.

Another rather empirical functional, which will become important in this work, was derived by Sharma et al. \cite{Sharma_al.2008}. They realized that both the exchange-only functionals as well as the M\"uller functional can be seen as instances of a more general functional, namely the $\alpha$ or Power functional, described by $f^{\alpha}(n_i,n_j)=(n_in_j)^\alpha$. As the M\"uller functional underestimates the correlation-energy, one expects to improve the results for values of $\alpha$ between 0.5 and 1. This assumption proved to be valid \cite{Lathiothakis_al.2009}, leading to an accurate description of the dissociation energy curve of H2 (a discussion of the performance of several DFT functionals for this problem can be found in \cite{Ayers_al.1998}).

These functionals were mainly constructed to describe small finite systems. However, because the correlation functional should be general, they should also work in the case of extended systems like the HEG. We will show the performance of some of the previously mentioned functionals when applied to the HEG later on in Figure \ref{fig.rdmft.ec-3d}. 

The BBC2/3 and PNOF0/1 functionals show a behaviour similar to the one shown by BBC1 and ML-SIC performs even worse than ML. 
From the functionals mentioned so far, only the $\alpha$ functional is capable of describing the correlation-energy over the whole range of densities considerably accurately.

To make some statements about whether or not a given functional reproduces an accurate momentum distribution, we will rely on the results by Gori-Giorgi and Ziesche \cite{Giorgi_Ziesche.2002}.  Three important properties of the momentum distribution derived from these results are listed in the following.
\begin{enumerate}
  \item [(1)] Nonzero occupation of all high momentum states \cite{Kimball.1975} which can be understood by the help of the electronic cusp condition \cite{Friesecke.2003,Baldsiefen_al_1.2012}.
  \item [(2)] A discontinuity and rather symmetrical behaviour of the momentum distribution at the Fermi level (as also suggested by Landau-Liquid theory).
  \item [(3)] Depletion of low momentum states.
\end{enumerate}
All but one functionals mentioned so far succeed in recovering property (1). This is due to the big value of the derivative $\frac{\partial E_{xc}[\gamma]}{\partial n_i}$ for $n_i\rightarrow0$. All but the ML/ML-SIC functionals show a divergence for this derivative which will then lead to a partial occupation of all states. The ML/ML-SIC functionals exhibit only a very big, but finite, derivative. Therefore, states of very high momentum are not to be expected to show partial occupation. When it comes to the description of property (2), only the BBC and the PNOF functionals recover a discontinuity at the Fermi  surface. However, the overall behaviour around the Fermi level is not symmetric or qualitatively resembling the Monte-Carlo results. Furthermore, the size of the discontinuity, created by the BBC functionals, shows a wrong behaviour when changing $r_s$. Although expected to decrease with decreasing density, it increases \cite{Lathiotakis_Helbig_Gross.2007}. Property (3), a qualitatively correct depletion of low momentum states, is not fulfilled by the investigated functionals. Only the BBC1, BBC2 and PNOF functionals show a small decrease of occupation, but not nearly as much as prevalent in the exact momentum distributions. 

We believe that the depletion of low momentum states is an important physical effect which will become increasingly important in the low density limit and should therefore be recovered by an \abref{RDMFT} functional. Motivated by this credo, we are going to design an appropriate functional in the following.

\subsubsection{BOW functional}\label{sec.rdmft.bow}
We want to construct an \abref{xc} functional which is capable of reproducing both occupation of high momentum states as well as depletion of low momentum ones. For the time being we do not focus on the discontinuity at the Fermi level.

\begin{figure}[t!]
  \begin{center}
    \psfrag{tx}[Bc][Bc][1][0]{$n_in_j$}
    \psfrag{ty}[Bc][Bc][1][0]{$f$}
    \psfrag{thf}[Bl][Bl][0.8][0]{$f^{HF}$}
    \psfrag{tb5}[Bl][Bl][0.8][0]{$f^{BOW}(0.5)$}
    \psfrag{ta5}[Bl][Bl][0.8][0]{$f^\alpha(0.5)$}
    \psfrag{tb7}[Bl][Bl][0.8][0]{$f^{BOW}(0.7)$}
    \psfrag{ta7}[Bl][Bl][0.8][0]{$f^\alpha(0.7)$}
    \includegraphics[width=\columnwidth]{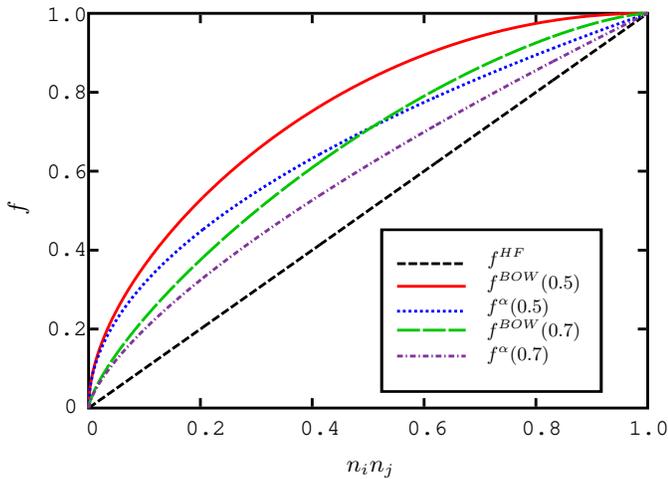}
    \caption{Occupation number dependence function $f$ for the BOW and $\alpha$ functionals for different values of the parameter $\alpha$. The Hartree-Fock function is reproduced by both functionals with $\alpha=1$.}\label{fig.f-bow-alpha}
  \end{center}
\end{figure}

\begin{figure}[b!]
  \begin{center}
    \psfrag{tx}[Bc][Bc][1][0]{$r_s(a.u.)$}
    \psfrag{ty}[Bc][Bc][1][0]{$e_c(Ha)$}
    \psfrag{mc}[Bl][Bl][0.8][0]{PWCA} 
    \psfrag{tv}[Bl][Bl][0.8][0]{BOW ($0.61$)} 
    \psfrag{bbc}[Bl][Bl][0.8][0]{BBC1} 
    \psfrag{tb2}[Bl][Bl][0.8][0]{BBC2} 
    \psfrag{cga}[Bl][Bl][0.8][0]{CGA}     
    \psfrag{ml}[Bl][Bl][0.8][0]{ML}   
    \psfrag{mue}[Bl][Bl][0.8][0]{M\"uller} 
    \psfrag{pow}[Bl][Bl][0.8][0]{$\alpha\ (0.55)$} 
    \psfrag{bg12}[Bl][Bl][0.8][0]{BOW (0.61)} 
    \includegraphics[width=\columnwidth]{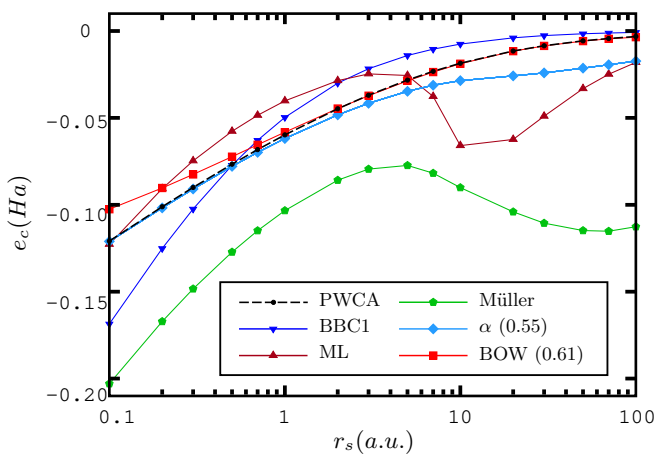}
    \caption{Correlation-energy of the three-dimensional electron gas using various correlation functionals from RDMFT. The black line denotes Monte-Carlo results in the \abref{PWCA} \cite{Perdew_Wang.1992} parametrization.}\label{fig.rdmft.ec-3d}
  \end{center}
\end{figure}

Considering the occupation of high momentum states, we will let us be guided by the success of the $\alpha$ functional and include a term of $(n_in_j)^\alpha$ in our functional. To achieve a depletion of low momentum states we then require our functional to have a vanishing derivative for $n_i=n_j=1$. In this way it is possible to reduce the occupation number of fully occupied orbitals without changing the energy. As the exchange contribution is negative, this excess charge can be used to lower the energy. The variational principle will therefore lead to a groundstate where the orbitals are never fully occupied. A possible choice of functional with vanishing derivative for full occupation could then be $f(n_i,n_j)=((n_in_j)^\alpha-\alpha n_in_j)/(1-\alpha)$. We found, however, that this choice underestimates the correlation-energy considerably. The partially occupied states are given too much influence on the energy. We therefore introduce a simple counter-term to decrease this effect. It incorporates the inverse of the $\alpha$ functional, leading to our final choice for the \abref{xc} functional.
\begin{multline}
  E^{BOW}_{xc}[\gamma;\alpha]=-\frac12\sum_{ij}f^{BOW}(n_i,n_j;\alpha)\\
  \int dxdx'w(x,x')\phi_i^*(x')\phi_i(x)\phi_j^*(x)\phi_j(x')
\end{multline}
\begin{multline}
  f^{BOW}(n_i,n_j;\alpha)=(n_in_j)^\alpha-\alpha n_in_j+\alpha\\
  -\alpha(1-n_i n_j)^{1/\alpha}
\end{multline}

As a neccessary property, the BOW-functional leads to the reproduction of the exchange-only functional for uncorrelated momentum distributions
\begin{align}
  f^{BOW}(n_i,n_j;\alpha)|_{n_in_j=0}&=0\\
  f^{BOW}(n_i,n_j;\alpha)|_{n_i=n_j=1}&=1.
\end{align}
Furthermore, we recover the exchange-only functional by choosing $\alpha=1$, i.e. $f^{BOW}(n_i,n_j;1)=f^x(n_i,n_j)$.
We can now use the parameter $\alpha$ to tune the influence of correlation. We show $f^{BOW}(n_i,n_j;\alpha)$ and the function from the $\alpha$-functional $f^\alpha(n_i,n_j;\alpha)=(n_in_j)^\alpha$ for several values of $\alpha$ in Figure \ref{fig.f-bow-alpha}. As we can see, a decrease in $\alpha$ leads to a bow-like shape which lead to our choice of name.

\begin{figure}[t!]
  \begin{center}
    \psfrag{mc}[Bl][Bl][0.8][0]{MC} 
    \psfrag{bg12}[Bl][Bl][0.8][0]{BOW $(0.66)$} 
    \psfrag{bbc}[Bl][Bl][0.8][0]{BBC1} 
    \psfrag{tb2}[Bl][Bl][0.8][0]{BBC2} 
    \psfrag{cga}[Bl][Bl][0.8][0]{CGA}     
    \psfrag{ml}[Bl][Bl][0.8][0]{ML}     
    \psfrag{mue}[Bl][Bl][0.8][0]{M\"uller} 
    \psfrag{pow}[Bl][Bl][0.8][0]{$\alpha (0.6)$} 
    \psfrag{tx}[Bc][Bc][1][0]{$r_s(a.u.)$}
    \psfrag{ty}[Bc][Bc][1][0]{$e_c(Ha)$}
    \includegraphics[width=\columnwidth]{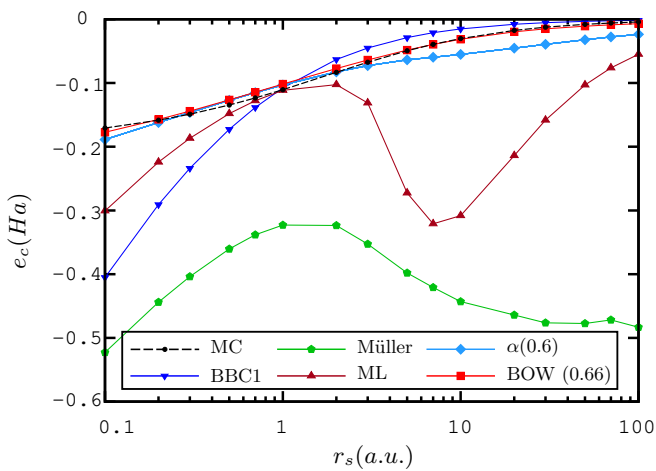}
  \end{center}
  \caption{Correlation-energy of the two-dimensional electron gas. The black line denotes Monte-Carlo results in the Attaccalite \cite{Attaccalite_al.2002} parametrization.}
  \label{fig.rdmft.ec-2d}
\end{figure}

The performance of the BOW functional when applied to the HEG in 3 and 2 dimensions in comparison to other prominent RDMFT functionals is shown in Figures \ref{fig.rdmft.ec-3d} and \ref{fig.rdmft.ec-2d}.

\begin{figure}[t!]
  \centering
  \psfrag{tx}[Bc][Bc][1][0]{$k/k_F$}
  \psfrag{ty}[Bc][Bc][1][0]{$n(k/k_F)$}
  \psfrag{tt1}[Bc][Bc][0.8][0]{$r_s=0.5$}
  \psfrag{tt2}[Bc][Bc][0.8][0]{$r_s=1$}
  \psfrag{tmc05}[Bl][Bl][0.8][0]{GZ} 
  \psfrag{tv05}[Bl][Bl][0.8][0]{BOW} 
  \psfrag{tmc1}[Bl][Bl][0.8][0]{GZ} 
  \psfrag{tv1}[Bl][Bl][0.8][0]{BOW} 
  \includegraphics[width=.8\columnwidth]{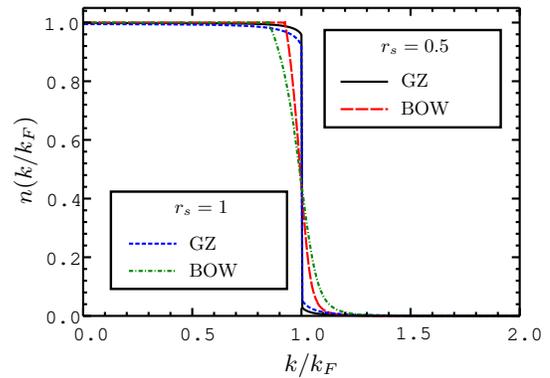} 
  \caption{Momentum distributions of the 3D-\abref{HEG} for $r_s\in\{0.5,1\}a.u.$ from the parametrization by Gori-Giorgi and Ziesche (GZ) \cite{Giorgi_Ziesche.2002} and from the \abref{BOW} functional for $\alpha=0.61$.}\label{fig.rdmft.mom-dist.05-1}
\end{figure}

\begin{figure}[t!]
  \psfrag{tx}[Bc][Bc][1][0]{$k/k_F$}
  \psfrag{ty}[Bc][Bc][1][0]{$n(k/k_F)$}
  \psfrag{tt1}[Bc][Bc][0.8][0]{$r_s=5$}
  \psfrag{tt2}[Bc][Bc][0.8][0]{$r_s=10$}
  \psfrag{tmc5}[Bl][Bl][0.8][0]{GZ} 
  \psfrag{tv5}[Bl][Bl][0.8][0]{BOW} 
  \psfrag{tmc10}[Bl][Bl][0.8][0]{GZ} 
  \psfrag{tv10}[Bl][Bl][0.8][0]{BOW} 
  \includegraphics[width=.8\columnwidth]{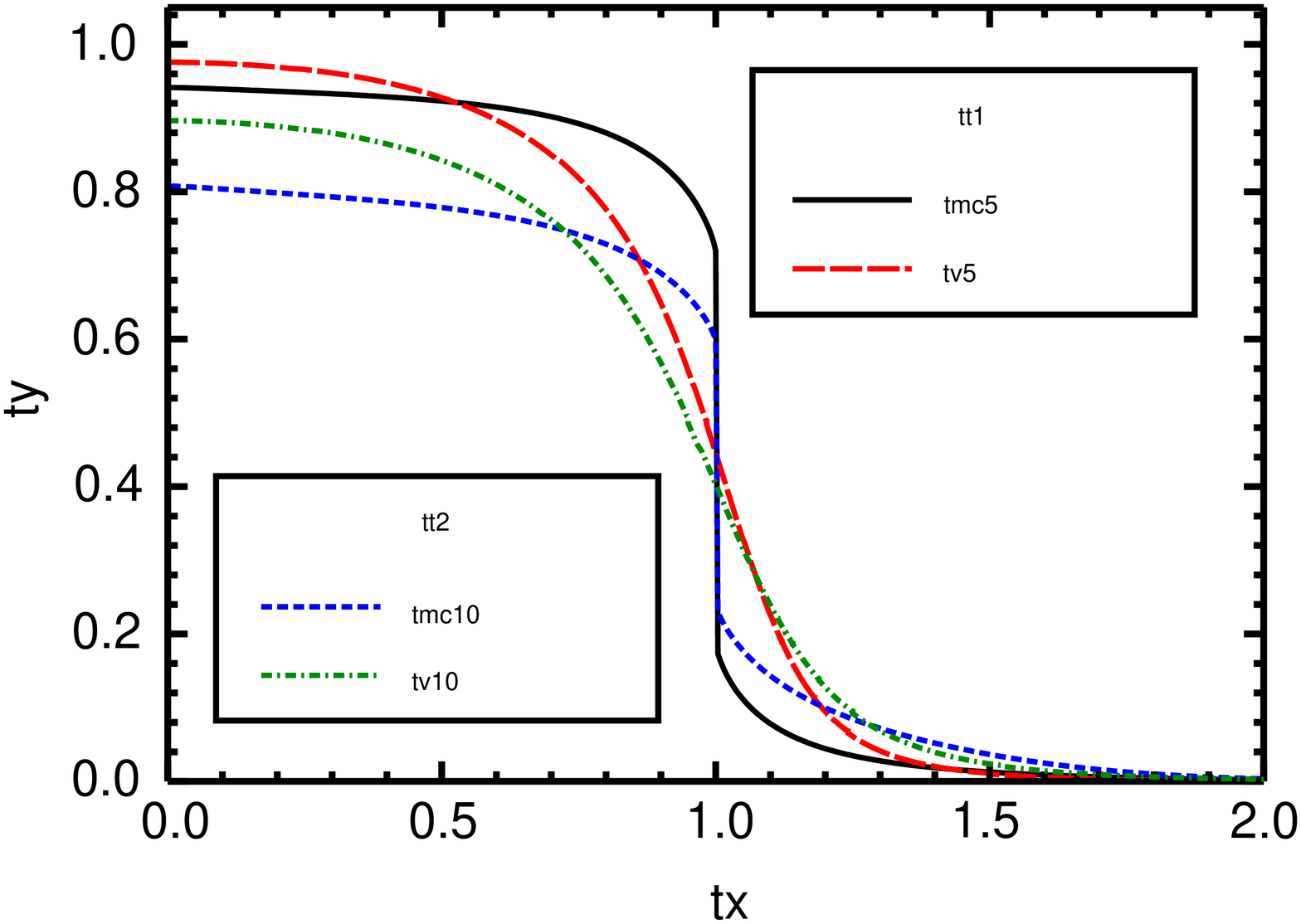} 
    \caption{Momentum distributions of the 3D-\abref{HEG} for $r_s\in\{5,10\}a.u.$ from the parametrization by Gori-Giorgi and Ziesche (GZ) \cite{Giorgi_Ziesche.2002} and from the \abref{BOW} functional for $\alpha=0.61$.}\label{fig.rdmft.mom-dist.5-10}
\end{figure}

Furthermore, we show several momentum distributions as resulting from the BOW functional in Figures \ref{fig.rdmft.mom-dist.05-1} and \ref{fig.rdmft.mom-dist.5-10}.

In summary, the \abref{BOW} functional shows remarkable success in describing the correlation-energy of the \abref{HEG} over a wide range of densities ($0.1<r_s<100$), including the range of metallic densities ($1<r_s<6$). It manages to qualitatively correctly describe the depletion of low momentum states, an exact property known from Monte-Carlo results.

Most of the functionals (M\"uller, BBC1/2/3,PNOF/0,ML,ML-SIC) currently on the market are spin-channel separable which from our findings in the previous section will lead to a qualitatively bad performance when applied to the spin polarized HEG. The $\alpha$ and \abref{BOW} functionals, on the other hand, offer a simple way out of the seperability dilemma because they incorporate a parameter. Making this parameter polarization-dependent makes the functional inseparable and offers an easy solution to the problem, given that the functionals also describe the partially polarized systems accurately. We have investigated the $\alpha$ and \abref{BOW} functionals for partially polarized systems and found a good agreement of the correlation energies for the respective optimal parameters which we show in Table \ref{tab.rdmft.pol-param}. However, although both the $\alpha$ as well as the \abref{BOW} functional with polarization-dependent parameter describe the correlation-energy qualitatively correctly, they fail to predict a quantum phase transition between unpolarized and polarized phases. This is due to the fact that a smaller coefficient in the low density limit in both functionals leads to a bigger contribution to the correlation-energy and therefore, by reproducing the energy for intermediate densities well, they favour the unpolarized phase for low density, where the phase transition should occur.

\setlength{\tabcolsep}{4pt}
\begin{table}[t!]
  \centering
  \begin{tabular}[t]{|x{9mm}||x{9mm}|x{9mm}|x{4mm}|x{9mm}||x{9mm}|x{9mm}|}
    \cline{1-3}\cline{5-7}
    \multicolumn{3}{|l|}{3D}&&\multicolumn{3}{|l|}{2D}\\
    \hhline{===~===}
    $\xi$&$\alpha$&BOW&&$\xi$&$\alpha$&BOW\\
    \hhline{===~===}
    0.0&0.56&0.61&&0.0&0.63&0.66\\
    \cline{1-3}\cline{5-7}
    0.1&0.56&0.61&&0.1&0.63&0.66\\
    \cline{1-3}\cline{5-7}
    0.2&0.56&0.61&&0.2&0.63&0.66\\
    \cline{1-3}\cline{5-7}
    0.3&0.57&0.62&&0.3&0.63&0.67\\
    \cline{1-3}\cline{5-7}
    0.4&0.57&0.62&&0.4&0.64&0.67\\
    \cline{1-3}\cline{5-7}
    0.5&0.57&0.63&&0.5&0.64&0.68\\
    \cline{1-3}\cline{5-7}
    0.6&0.58&0.63&&0.6&0.65&0.69\\
    \cline{1-3}\cline{5-7}
    0.7&0.59&0.64&&0.7&0.66&0.71\\
    \cline{1-3}\cline{5-7}
    0.8&0.61&0.65&&0.8&0.67&0.73\\
    \cline{1-3}\cline{5-7}
    0.9&0.63&0.67&&0.9&0.70&0.76\\
    \cline{1-3}\cline{5-7}
    1.0&0.66&0.69&&1.0&0.74&0.80\\
    \cline{1-3}\cline{5-7}
  \end{tabular}
  \caption{Best parameters to describe the correlation-energy of a \abref{HEG} in three and two dimensions in the range $1<r_s<100$ for the $\alpha$ and \abref{BOW} functionals.}\label{tab.rdmft.pol-param}
\end{table}
\setlength{\tabcolsep}{6pt}

\begin{table}[b!]
  \centering
  \begin{tabular}[t]{|c|x{9mm}|x{9mm}|x{9mm}|}
    \hline
    &$\alpha^P$&$\alpha^U$&$c$\\
    \hline
    \abref{BOW-TIE}&0.70&2.0&0.19\\
    \hline  
  \end{tabular}
  \caption{Optimal parameters for the TIE-version of the \abref{BOW} \abref{xc} functional as defined in Eq. \eq{eq.rdmft.xc-prop}.}\label{tab.rdmft.pol-param.x}
\end{table}

This is a backlash because the quantum phase transition is a physical property of big interest. We will now propose a rather phenomenological way to recover the quantum phase transition at zero temperature. From the good agreement of the \abref{BOW} functional results with the Monte-Carlo results for fully polarized configurations, we deduce that one can, at least to some extent, describe the equal-spin-channel correlation effects by employing the exchange integral. We therefore propose to use a similar approach to include the opposite-spin-channel contributions additionally. Our expression for this ``trans-channel interaction energy'' (TIE) modification reads with an explicit mentioning of the spin index
\begin{multline}
  E^{BOW-TIE}_{xc}[\gamma]=-\frac12\sum_{ij\sigma}f^{BOW}(n_{i\sigma},n_{j\sigma};\alpha^P)K(i,j)-\\
  c\sum_{ij}\left(\left(n_{i\uparrow}n_{j\downarrow}\right)^{\alpha^U}\left(1-n_{i\uparrow}n_{j\downarrow}\right)^{\alpha^U}\right)K(i,j)\label{eq.rdmft.xc-prop}.
\end{multline}
$K(i,j)$ represents the exchange integral, corresponding to the \abref{NO}s $\phi_i$ and $\phi_j$ and $\alpha^P$ stand for the best parameter for the description of the fully polarized \abref{HEG}. The second term in Eq. \ref{eq.rdmft.xc-prop} vanishes for a spin-polarized system and contributes increasingly with decreasing polarization. The coefficients $\alpha^U$ and $c$ are fitted to reproduce a critical density closer to the Monte-Carlo result while maintaining the good overall accuracy of the correlation-energy for different spin polarizations. The resulting parameters are shown in Table \ref{tab.rdmft.pol-param.x} and lead to an instantaneous phase transition at a critical density of $r_c\approx 28a.u.$.

We would like to emphasize again that we did not deduce this opposite-spin-channel contribution from higher principles but rather postulated it to create a model functional which reproduces the critical density of the \abref{HEG} more accurately. With a different choice for the inter-spin channel correlation energy, i.e. one which favours partially polarized configurations more strongly, one might be able to get rid of the instantaneous transition between unpolarized and polarized phases and reproduce a qualitatively correct continuous quantum phase transition.

This concludes our investigation of zero-temperature RDMFT and we are now able to investigate the effect of temperature in the framework of FT-RDMFT

\subsection{Finite Temperature RDMFT}

\subsubsection{DFT-LSDA correlation}
Firstly, we are going to employ the correlation-energy of the electron gas as parametrized in the PWCA approximation. Remembering the conceptual difference between RDMFT and DFT, i.e. the knowledge of the exact kinetic energy functional, one might be tempted to remove the kinetic contribution from the PWCA-functional before employing it as a correlation-energy functional in RDMFT. This procedure, however, would be incorrect as will be explained in the following.

\begin{figure}[t!]
  \centering
  \psfrag{tx}[bc][bc][1.0][0]{$T(K)$}
  \psfrag{ty}[bc][bc][1.0][0]{$r_s(a.u.)$}
  \psfrag{tf}[bc][bc][1.0][0]{$T_F$}
  \psfrag{ttc}[bc][bc][1.0][0]{$(r_c^T,T_c)$}
  \psfrag{trc}[bc][bc][1.0][0]{$r_c$}
  \includegraphics[width=\columnwidth]{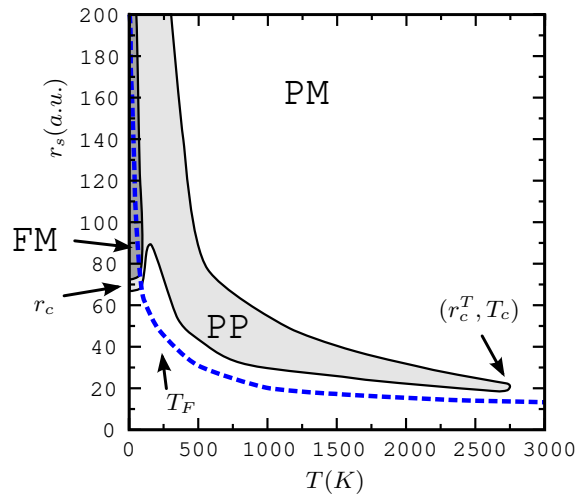}
  \caption{Collinear phase diagram of the \abref{HEG} for the \abref{DFT}-\abref{LSDA} functional. FM: ferromagnetic phase, PM: paramagnetic phase, PP: partially polarized phase. The dotted blue line denotes the Fermi temperature $T_F$.}
  \label{fig.applications.pd-lsda}
\end{figure}

Based on our previous discussion of the intrinsic problems of the utilization of DFT functionals in RDMFT we know that the minimizing momentum distribution will be ``uncorrelated'', i.e. will reproduce the Hartree-Fock solution which becomes a step function $n^0(\tb k)$ in the $T\rightarrow0$ limit. If we removed the kinetic contribution the correlation-energy functional would be given by the interaction contribution $W^{PWCA}_c$ alone and the free energy at density $\rho$ would read
\begin{align}
  F&\stackrel{T\rightarrow0}{=}\Omega_k[n^0(\tb k)]+\Omega_x[n^0(\tb k)]+W^{PWCA}_c(\rho)
\end{align}
Because the momentum distribution is a step function, this expression reproduces exactly the Monte-Carlo results without the kinetic contribution. This poses a problem because removing the kinetic contribution from a DFT-correlation functional yields an overall bad approximation. A straightforward remedy of this problem is the re-inclusion of the kinetic contribution to accomodate for the ``uncorrelated'' nature of the momentum distribution. Cashing in on our previous discussion we are therefore able to construct a FT-RDMFT functional which reproduces the exact Monte-Carlo results in the $T\rightarrow0$ limit. The resulting phase diagram is shown in Figure \ref{fig.applications.pd-lsda}.

By construction, we recover the Monte-Carlo results in the PWCA parametrization for $T\rightarrow0$. This leads to the reproduction of the continuous quantum phase transition at the correct critical Wigner-Seitz radius of $r_c\approx70a.u.$. In the radius range of $20a.u.$ to $70a.u.$ with an increase of temperature we encounter a continuous phase transition to a partially polarized state. A further increase of temperature then leads to a continuous phase transition back to the paramagnetic configuration. Considering the density range defined by $70a.u.<r_s<90a.u.$, the situation becomes more complicated. Starting from the polarized configuration at zero temperature we encounter first a phase transition to a paramagnetic state, then another phase transition to a partially polarized state, and then a final transition back to the paramagnetic configuration. We attribute the appearance of the big partially polarized phase for temperatures above $T_F$ to the fact that the \abref{1RDM} is ''uncorrelated`` and therefore the noninteracting entropy functional underestimates the real entropy. It would be interesting to investigate, how a FT-\abref{LSDA} from \abref{DFT} would perform in the \abref{FT-RDMFT} framework, but as mentioned before, the necessary Monte-Carlo results do not exist so far.  

We will now turn to the investigation of a true RDMFT correlation functional.

\subsubsection{BOW-TIE correlation}

\begin{figure}[t!]
  \centering
  \psfrag{tx}[bc][bc][1.0][0]{$T(K)$}
  \psfrag{ty}[bc][bc][1.0][0]{$r_s(a.u.)$}
  \psfrag{ttf}[bc][bc][1.0][0]{$T_F$}
  \psfrag{ttc}[bc][bc][1.0][0]{$(r_c^T,T_c)$}
  \psfrag{trc}[bc][bc][1.0][0]{$r_c$}
  \includegraphics[width=\columnwidth]{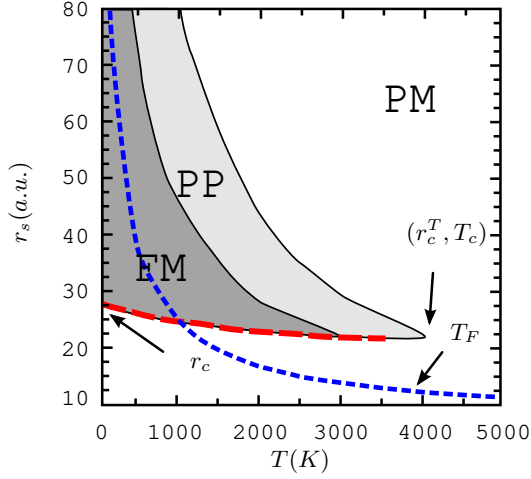}
  \caption{Collinear phase diagram of the \abref{HEG} for the BOW-TIE functional. FM: ferromagnetic phase, PM: paramagnetic phase, PP: partially polarized phase. The thick dashed red line denotes an instantaneous phase transitions. The dotted blue line denotes the Fermi temperature $T_F$.}
  \label{fig.applications.bow}
\end{figure}

Following from our considerations regarding zero temperature RDMFT we employed the BOW-TIE functional as defined in Eq \eq{eq.rdmft.xc-prop} with the coefficients from Table \ref{tab.rdmft.pol-param.x}. The resulting phase diagram is shown in Figure \ref{fig.applications.bow}. The critical density of the instantaneous transition increases with increasing temperature and the critical temperature is $T\approx 4T_F$.

Apparently, the temperature independent BOW-TIE functional underestimates the effect of correlation as induced by temperature. To remedy this problem we will therefore in the following consider explicitly temperature-dependent functionals. We will again start with a DFT functional, namely the FT-RPA functional and will then propose a temperature dependent true FT-RDMFT correlation functional.

\subsubsection{FT-DFT-RPA correlation}
\begin{figure}[t!]
  \centering
  \psfrag{tx}[bc][bc][1.0][0]{$T(K)$}
  \psfrag{ty}[bc][bc][1.0][0]{$r_s(a.u.)$}
  \psfrag{tf}[bc][bc][1.0][0]{$T_F$}
  \psfrag{ttc}[bc][bc][1.0][0]{$(r_c^T,T_c)$}
  \psfrag{trc}[bc][bc][1.0][0]{$r_c$}
  \includegraphics[width=\columnwidth]{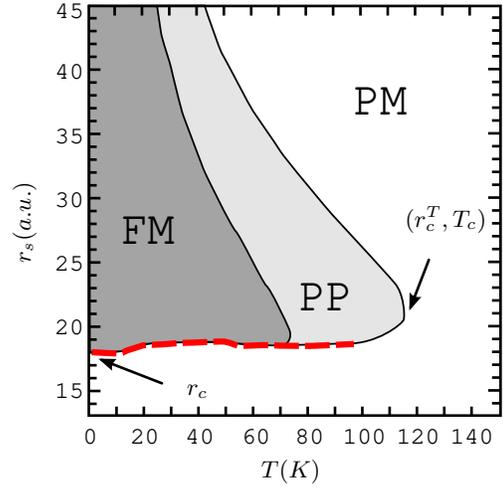}
  \caption[\abref{HEG} phase diagram from \abref{FT-RDMFT} with \abref{FT-DFT}-\abref{RPA} correlation]{Collinear phase diagram of the \abref{HEG} for the \abref{FT-DFT}-\abref{RPA} functional. The thick dashed red line denotes first-order phase transitions. $T_F$ is too big to fit in the graph.}
  \label{fig.applications.pd-ftdft-rpa}
\end{figure}

The correlation contribution to the grand potential in the FT-RPA is defined as
\begin{multline}
  \Omega_r=\frac{1}{2\beta}\int\frac{d^3q}{(2\pi)^3}\sum_{a=-\infty}^{\infty}\\
  \left\{\ln\left(1-W(\tb q)\chi(\tb q,\nu_a)\right)+W(\tb q)\chi(\tb q,\nu_a)\right\}\label{eq.gpr},
\end{multline}
where the frequencies $\nu_a$ are given by $\nu_a=\frac{2\pi a}{\beta}$ and the polarization propagator is defined as
\begin{align}
  \chi(\tb q,\nu_a)&=\chi_{\uparrow}(\tb q,\nu_a)+\chi_{\downarrow}(\tb q,\nu_a)\label{eq.gpr.chi}\\
  \chi_\sigma(\tb q,\nu_a)&=-\int\frac{d^3k}{(2\pi)^3}\frac{n_\sigma(\tb k+\tb q)-n_\sigma(\tb k)}{\tb i\nu_a-(\veps_\sigma(\tb k+\tb q)-\veps_\sigma(\tb k))}\label{eq.gpr.chi-sigma}
\end{align}
In the case of DFT, the momentum distribution $n(\tb k)$ is given by the noninteracting Fermi-Dirac distribution. Results for the unpolarized HEG can be found in Refs. \cite{Gupta_Rajagopal_2.1980,Gupta_Rajagopal.1982}. We calculated the RPA correlation grand potential for arbitrary polarization and temperature and show the resulting phase diagram in Figure \ref{fig.applications.pd-ftdft-rpa}. As the \abref{1RDM} is ''uncorrelated'',  at zero temperature the momentum distribution will be the same as in the \abref{FT-MBPT} treatment, namely a step function. Therefore, we recover the critical Wigner-Seitz radius from zero-temperature RPA of $r_c\approx18a.u.$. Compared to our general expectations the phase diagram only fails to correctly predict the critical temperature $T_c$ is an order of magnitude smaller than the Fermi temperature $T_F\approx1300K$ of the corresponding critical radius $r_c^T\approx21a.u.$. We therefore deduce that the FT-DFT-RPA functional employed in the FT-RDMFT framweork manages to increase the critical Wigner-Seitz radius but overestimates the effect of temperature.

The three functionals investigated so far were either intrinsically temperature independent or did just depend on the density alone. As a final approximation we will therefore now derive a correlation functional in FT-RDMFT which will be both explicitly temperature dependent and it will be a functional of the full momentum distribution.

\subsubsection{FT-RDMFT-RPA functional}
We will utilize the perturbative expansion method as developed in Part I of this work \cite{Baldsiefen_al_1.2012}. We choose the subset of diagrams which refer to the RPA diagrams in FT-MBPT which are defined in Eqs. \eq{eq.gpr} - \eq{eq.gpr.chi-sigma}. Whereas the contributions $\Omega_k,\Omega_{ext},S_0$, and $\Omega_x$ only depend on the momentum distribution directly, $\Omega_r$ (and all higher-order diagrams which exhibit more independent momenta than noninteracting Green's functions) also depend explicitly on the Kohn-Sham energies. This becomes a problem in FT-RDMFT as we will show in the following. Using the fundamental one-to-one correspondence $\rho\leftrightarrow v_{KS}$, we see that $\rho$ determines $v_{KS}$ which determines $\veps(k)$. Hence the $\veps(k)$ are functionals of $\rho$ and $\Omega_r$ can be viewed as an implicit functional of $\veps(k)$. To get a rough idea of the behaviour of this functional we parametrize the Kohn-Sham energies in the following way.
\begin{align}
  \veps(m;\tb k)&=\frac{k^2}{2m}\label{eq.model.disp},
\end{align}
where we choose $m$ to be a variational parameter which can be interpreted as an effective mass. It will now be favourable to work with a reduced temperature $t=T/T_F$ because it was shown \cite{Gupta_Rajagopal_1.1980} that for a noninteracting system the fugacity $\alpha=\beta\mu$ only depends on $t$ alone. From Eq. \eq{eq.ks.ee} we see that a dispersion relation as in Eq. \eq{eq.model.disp} leads to a simple dependence of the fugacity on $m$, namely $\alpha(t;m)=\alpha(mt)$, and the momentum distribution becomes
\begin{align}
  n(t,m;\tb k)&=\frac{1}{1+e^{\frac{k^2}{mt}-\alpha(mt)}}=n(mt,1;\tb k).\label{eq.model.nk}
\end{align}
The general effect of $m$ on the momentum distribution is then that a decrease of $m$ leads to a smoother momentum distribution. We are therefore able to investigate qualitatively the behaviour of \abref{FT-RDMFT} functionals under a change in $n(\tb k)$. Eq. \eq{eq.model.nk} tells us that all functionals which just depend on the momentum distribution show the same simple dependence on $m$.
\begin{align}
  \Omega_{k}(t,m)&=\Omega_{k}(mt,1)\\
  \Omega_{ext}(t,m)&=\Omega_{ext}(mt,1)\\
  N(t,m)&=N(mt,1)\\
  S_{0}(t,m)&=S_{0}(mt,1)\\
  \Omega_{x}(t,m)&=\Omega_{x}(mt,1)\\
  F(t,m)&=E_{k}(mt,1)+\Omega_{ext}(mt,1)-\nn\\
  &\hspace*{14mm}1/\beta S_{0}(mt,1)+\Omega_{x}(mt,1)
\end{align}
It has to be noted that because of the temperature prefactor of $S_0$, $F(t,m)\neq F(mt,1)$. The polarization propagator in \abref{RPA}, on the other hand, shows a different behaviour because of the explicit dependence on $\veps(m;\tb k)$.
\begin{align}
  \chi_\sigma(t,m;q,\nu_a)&=m\chi_\sigma(mt,1;q,\nu_a)
\end{align}
Let us now consider the limit of $t\rightarrow0$ while keeping the product $mt$, and therefore the momentum distribution, fixed, i.e. $t\rightarrow0,m\propto1/t$. All contributions up to first order in the interaction stay invariant under this transformation. The free energy will change, because of the prefactor $1/\beta$ in front of $S_0$. However, under the assumption that we started from a finite entropy this free energy change will be finite. The polarization propagator on the other hand will diverge because it exhibits a prefactor of $m$ which leads to a divergence of $\Omega_r$.

Taking the previous considerations into account, we come to the conclusion that a straightforward inclusion of subsets of higher-order diagrams in the methodology derived in \cite{Baldsiefen_al_1.2012} will most likely lead to ill-behaved functionals, yielding wrong energies and momentum distributions. It has to be noted that this does not disprove the validity of the perturbation expansion of $\Omega[\gamma]$. It only shows that the utilization of a subset of diagrams in a variational scheme carries the danger of leading to a variational collapse. This fact is to be attributed to the total freedom of choice for $\veps(\tb k)$ by the inclusion of nonlocal potentials, a fact also recently pointed out in the context of GW \cite{Ismail-Beigi.2010}. As a general recipe for avoiding the problem of variational collapse one might model the perturbative expressions by approximations using the momentum distributions alone. We are going to investigate several first steps in this direction in the following.

As we have seen before, the divergence in $\Omega_r$ stems from the fact that the eigenenergies $\veps_\sigma(\tb k)$ appear exlicitely in the definition of $\chi_\sigma$. A first guess to remedy this problem is to fix these eigenenergies to the noninteracting values.
\begin{align}
  \veps_\sigma(\tb k)&=\veps^0_\sigma(\tb k)=\frac{k^2}{2}
\end{align}
We have implemented this correlation functional and found that it yields qualitatively incorrect results. The correlation functional in this approximation favours less washed out momentum distributions. 

This drawback lets us formulate a functional resting on exact properties of the polarization propagator. As a first step in our approximation process we will model $\chi(\tb q,\nu_a)$ as being frequency independent, much in the spirit of the COHSEX approximation \cite{Hedin.1965}. Using the relation
\begin{align}
  \frac{\partial n(\tb k)}{\partial \veps(\tb k)}&=-\beta n(\tb k)(1-n(\tb k)),
\end{align}
the $\tb q\rightarrow 0$ limit of the kernel of $\chi(\tb q,0)$ can be calculated up to second order to give
\begin{multline}
  \frac{n(\tb k+\tb q)-n(\tb k)}{\veps(\tb k+\tb q)-\veps(\tb k)}=\frac{\partial n(\tb k)}{\partial \veps(\tb k)}\\
  \left(1+\frac{\beta(n(\tb k)-\frac12)\sum_{i,j}q_iq_j\frac{\partial\veps(\tb k)}{\partial k_i}\frac{\partial \veps(\tb k)}{\partial k_j}}{\sum_iq_i\frac{\partial\veps(\tb k)}{\partial k_i}}\right).\label{eq.frac}
\end{multline}
We assume the ONs and eigenenergies to be point symmetrical around the origin, i.e. $n(\tb k)=n(-\tb k)$ and $\veps(\tb k)=\veps(-\tb k)$. Accordingly, the second term in Eq. \eq{eq.frac} vanishes in the integration over $\tb k$. The polarization propagator in the limit $\tb q\rightarrow0$ therefore becomes
\begin{align}
  \chi_\sigma(\tb q)&\stackrel{q\rightarrow 0}{\longrightarrow}-\beta\int\frac{d^3k}{(2\pi)^3}n_\sigma(\tb k)(1-n_\sigma(\tb k))+O(q^2)
\end{align}

We now need to describe the momentum dependence for big $\tb q$. Starting from the main expression for the RPA correlation grand potential Eq. \eq{eq.gpr} we use the spatial isotropy of the interacion and write
\begin{align}
  \Omega_r&=\frac{1}{\beta}(2\pi)^{-2}\int dq q^2\left\{\ln\left(1-W(\tb q)\chi(\tb q)\right)+W(\tb q)\chi(\tb q)\right\}\label{eq.gpr.q0}.
\end{align}
It is now possible to estimate the behaviour of $\Omega_r$. If $W(\tb q)\chi(\tb q)$ is big, the logarithm in the kernel of $\Omega_r$ becomes negligible. The main object defining whether or not $W(\tb q)\chi(\tb q)$ is big or small is $\beta/(q^2)$. We therefore split the integral in Eq. \eq{eq.gpr.q0} in two parts. One wherein $q<\widetilde q=\sqrt{\beta}$ and one wherein $q>\widetilde q$.
\begin{align}
  \Omega_r&\approx\Omega_r^{0}+\Omega_r^{\infty}\\
  \Omega_r^0&=\frac{1}{\beta}(2\pi)^{-2}\int_0^{\widetilde{q}} dq q^2(W(\tb q)\chi(\tb q))\\
  &=\frac{4\pi}{\beta}(2\pi)^{-2}\int_0^{\sqrt{\beta}} dq \chi(\tb q))\label{eq.gpr0}\\
  \Omega_r^{\infty}&=-\frac{1}{\beta}(2\pi)^{-2}\int_{\sqrt{\beta}}^\infty dq q^2\frac{(W(\tb q)\chi(\tb q))^2}{2}
\end{align}
It can be deduced from Eq. \eq{eq.gpr0} that in the limit $T\rightarrow0$, i.e. $\beta\rightarrow\infty$, $\Omega_r^0$ diverges unless $\chi(\tb q)$ behaves like
\begin{align}
  \chi(\tb q)&\stackrel{q\rightarrow \infty}{\longrightarrow}\propto q^{\kappa}\\
  \kappa&\leq-2;
\end{align}

We incorporate the findings from the investigation of the exact limits of $\chi$ in the following approximate $\kappa$-functional
\begin{align}
  \chi^{\kappa}_\sigma[n(\tb k)](\tb q)&=-\frac{\beta}{1+q^2/k_F^2}\int\frac{d^3k}{(2\pi)^3}n_\sigma(\tb k)(1-n_\sigma(\tb k))\label{eq.chi.v1_0}\\
  &=-\frac{\beta}{1+(q/k_F)^\kappa}\left(\rho-\langle\rho^2\rangle\right),
\end{align}
where
\begin{align}
  \langle\rho^2\rangle&=\int\frac{d^3k}{(2\pi)^3}n^2_\sigma(\tb k)
\end{align}
The prefactor of $\beta$ ensures that $\Omega_r$ stays nonvanishing in the limit $\beta\rightarrow\infty$.
The occurrence of $\left(\rho-\langle\rho^2\rangle\right)$ in this approximation shows that $\chi^{\kappa}_\sigma[n(\tb k)](\tb q)$ will favour a more washed out momentum distribution.

\begin{figure}[t!]
  \begin{center}
    \psfrag{tx}[Bc][Bc][1.0][0]{$r_s(a.u.)$}
    \psfrag{ty}[Bc][Bc][1.0][0]{$e_c(Ha)$}
    \psfrag{pwca}[Bl][Bl][0.8][0]{PWCA} 
    \psfrag{2.9}[Bl][Bl][0.8][0]{$\kappa=2.9$} 
    \psfrag{2.7}[Bl][Bl][0.8][0]{$\kappa=2.7$} 
    \psfrag{2.5}[Bl][Bl][0.8][0]{$\kappa=2.5$} 
    \psfrag{2.3}[Bl][Bl][0.8][0]{$\kappa=2.3$} 
    \includegraphics[width=\columnwidth]{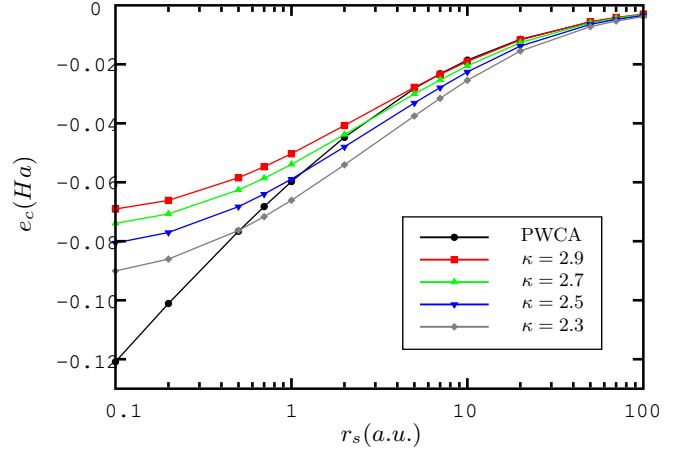}
    \caption{Correlation-energy of the three-dimensional electron gas for the $\kappa$-functional for various values of $\kappa$. The black line denotes Monte-Carlo results in the \abref{PWCA} \cite{Perdew_Wang.1992} parametrization.}\label{fig.rdmft.ec-rpa-kappa}
  \end{center}
\end{figure}

\begin{figure}[b!]
  \centering
  \psfrag{tx}[Bc][Bc][1][0]{$k/k_F$}
  \psfrag{ty}[Bc][Bc][1][0]{$n(k/k_F)$}
  \psfrag{tmc1}[Bl][Bl][0.8][0]{$r_s=1:$ GZ} 
  \psfrag{tmc5}[Bl][Bl][0.8][0]{$r_s=5:$ GZ} 
  \psfrag{tmc10}[Bl][Bl][0.8][0]{$r_s=10:$ GZ} 
  \psfrag{tk1}[Bl][Bl][0.8][0]{$r_s=1:\kappa$} 
  \psfrag{tk5}[Bl][Bl][0.8][0]{$r_s=5:\kappa$} 
  \psfrag{tk10}[Bl][Bl][0.8][0]{$r_s=10:\kappa$} 
  \includegraphics[width=.8\columnwidth]{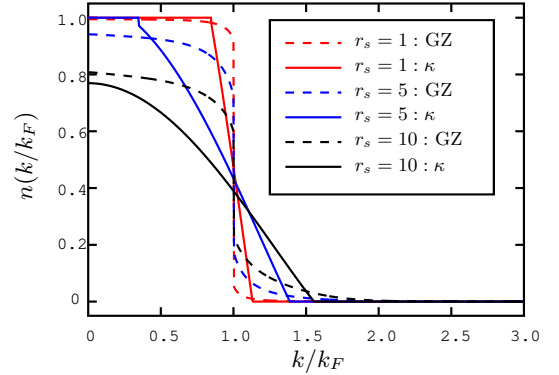} 
  \caption{Momentum distributions of the \abref{HEG} for $r_s\in\{1,5,10\}a.u.$ from the parametrization by Gori-Giorgi and Ziesche (GZ) \cite{Giorgi_Ziesche.2002} and from the $\kappa$-functional for $\kappa=2.9$.}\label{fig.rdmft.mom-dist.rpa}
\end{figure}

We calculated the correlation-energy in the zero-temperature limit for several values of $\kappa$ and show the results in Figure \ref{fig.rdmft.ec-rpa-kappa}. The momentum distributions resulting from $\Omega_r^{\kappa}[\gamma]$ for $\kappa=2.9$ are shown in Figure \ref{fig.rdmft.mom-dist.rpa}.

\setlength{\tabcolsep}{4pt}
\begin{table}[b!]
  \centering
  \begin{tabular}[t]{|x{9mm}|x{9mm}|x{9mm}|x{9mm}|x{9mm}|x{9mm}|x{9mm}|}
    \hline
    \multicolumn{7}{|l|}{$\kappa$-functional}\\
    \hline\hline
    $\xi$&0.0&0.1&0.2&0.3&0.4&0.5\\
    \hline
    $\kappa$&2.9&2.9&2.9&3.0&3.0&3.0\\
    \hline\hline
    $\xi$&0.6&0.7&0.8&0.9&1.0&\\
    \hline
    $\kappa$&3.1&3.1&3.1&3.2&3.2&\\
    \hline
  \end{tabular}
  \caption{Best parameters to describe the correlation-energy of a \abref{HEG} in three dimensions in the range $1<r_s<100$ for the $\kappa$-functional.}\label{tab.ft-rdmft.pol-param}
\end{table}
\setlength{\tabcolsep}{6pt}

Unfortunately, the $\kappa$-functional becomes spin-channel seperable in the $T\rightarrow0$ limit. As in our discussion of zero temperature RDMFT we can make the parameter polarization dependent to describe the partially polarized HEG correctly (see Table \ref{tab.ft-rdmft.pol-param}). However, for the same reasons as in the previous discussions the descreasing parameter for decreasing polarization leads to a favourisation of the unpolarized phase in the low density limit and therefore to the absence of a phase transition. We therefore propose a modification similar to the BOW-TIE functional as in Eq. \eq{eq.rdmft.xc-prop} to recover a favourisation of the ferromagnetic phase for low densities.
\begin{multline}
  \chi^{\kappa-TIE}_\sigma[n(\tb k)](\tb q)=-\chi^{\kappa}_\sigma[n(\tb k)](\tb q)-\\
  c\frac{\beta}{1+q^2/k_F^2}\int\frac{d^3k}{(2\pi)^3}n_\uparrow(\tb k)n_\downarrow(\tb k)(1-n_\uparrow(\tb k)n_\downarrow(\tb k))
\end{multline}
The parameters which lead to an accurate description of correlation energies for arbitrary polarization over the whole range of densities and a decrease in the critical density are shown in Table \ref{tab.ft-ft-rdmft.pol-param.x}.

\begin{table}[t!]
  \centering
  \begin{tabular}[t]{|c|x{9mm}|x{9mm}|}
    \hline
    &$\kappa$&$c$\\
    \hline
    \abref{$\kappa$-TIE}&3.2&0.24\\
    \hline  
  \end{tabular}
  \caption{Best parameters for the TIE-version of the $\kappa$-functional as defined in Eq. \eq{eq.chi.v1_0}.}\label{tab.ft-ft-rdmft.pol-param.x}
\end{table}

\begin{figure}[t!]
  \centering
  \psfrag{tx}[bc][bc][1.0][0]{$T(K)$}
  \psfrag{ty}[bc][bc][1.0][0]{$r_s(a.u.)$}
  \psfrag{ttf}[bc][bc][1.0][0]{$T_F$}
  \psfrag{ttc}[bc][bc][1.0][0]{$(r_c^T,T_c)$}
  \psfrag{trc}[bc][bc][1.0][0]{$r_c$}
  \includegraphics[width=\columnwidth]{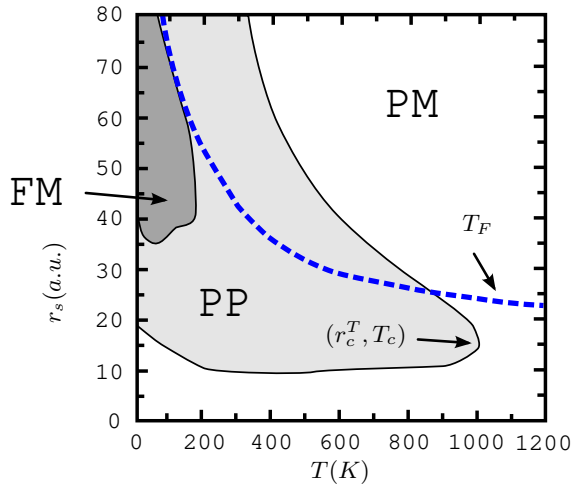}
  \caption{Collinear phase diagram of the \abref{HEG} for the $\kappa$-TIE functional. There are no first-order phase transitions. The dotted blue line denotes the Fermi temperature $T_F$.}
  \label{fig.applications.pd-ftrdmft-rpa}
\end{figure}

We conclude our work by showing the the phase diagram as resulting from the $\kappa$-TIE functional with this choice of parameters in Figure \ref{fig.applications.pd-ftrdmft-rpa}. We see that the unphysical instantaneous quantum phase transition disappears and an increase in temperature slightly favours a partially polarized configuration over a paramagnetic one. In summary, the $\kappa$-TIE functional constitutes an intrinsically temperature dependent true FT-RDMFT functional which describes the correlation-energy of a HEG at zero temperature with arbitrary polarization accurately, yields a qualtitatively improved momentum distribution, as compared to previously used RDMFT functionals, and leads to a reasonable magnetic collinear phase diagram.

\section{Conclusions}
In this work we have investigated the possibility of the description of correlation effects in a HEG in grand canonical equilibrium in the newly introduced framework of FT-RDMFT. We have first focussed on the zero temperature situation and have shown that a certain class of functionals, namely spin-channel seperable ones, will inevitably fail to describe the HEG for arbitrary polarization correctly. Using this knowledge we have then proposed a new functional, called BOW, which reproduces the correlation energies of a HEG at 3 or 2 dimensions for arbitrary polarization to unprecedented accuracy. It furthermore manages to describe the qualitative behaviour of the momentum distribution correctly.

We have then turned to the description of the HEG at finite temperature and have shown that correlation functionals from DFT will lead to ``uncorrelated'' momentum distributions. Nevertheless we have utilized the zero temperature PWCA functional as well as the FT-RPA functional in FT-RDMFT and have investigated the corresponding phase diagrams. 

Finally, we have investigated two truly momentum distribution dependent functionals. Firstly the BOW functional and secondly a newly proposed $\kappa$-functional which relies on exact limits of the RPA polarization propagator. 

This work being the first on the description of correlation in FT-RDMFT there are several possible propositions for further research and improvements. Above all the need for Monte-Carlo calculations for the HEG subject to nonlocal external potentials and/or finite temperature should be emphasized to get on the one hand a reliable measure for the accuracy of approximations and on the other hand the possibility of a real local-RDM-approximation. Further progress could be made by focussing on the correct prediction of the discontinuity of the momentum distribution at the Fermi energy, possibly by the different treatment of states below and above the Fermi surface.

In closing we would like to point out that the problem of the description of different magnetic phases of the HEG lies at the core of the problem of an accurate description of temperature dependent phases of real solids via a functional theory. However, the accurate description of the low density regime in which the HEG undergoes a phase transition is complicated by the fact that the energy differences between distinct phases become very small over a wide range of densities. Real systems, on the other hand, show phase transitions at much higher densities where the energy differences change strongly with respect to the density. The BOW and $\kappa$-functionals were designed to reproduce the Monte-Carlo results for different phases over the full range of densities, including the physical ones. We therefore expect that they are suitable candidates for the description of correlation effects in real solids which shall be a task for the future.


\begin{thebibliography}{45}
\expandafter\ifx\csname natexlab\endcsname\relax\def\natexlab#1{#1}\fi
\expandafter\ifx\csname bibnamefont\endcsname\relax
  \def\bibnamefont#1{#1}\fi
\expandafter\ifx\csname bibfnamefont\endcsname\relax
  \def\bibfnamefont#1{#1}\fi
\expandafter\ifx\csname citenamefont\endcsname\relax
  \def\citenamefont#1{#1}\fi
\expandafter\ifx\csname url\endcsname\relax
  \def\url#1{\texttt{#1}}\fi
\expandafter\ifx\csname urlprefix\endcsname\relax\def\urlprefix{URL }\fi
\providecommand{\bibinfo}[2]{#2}
\providecommand{\eprint}[2][]{\url{#2}}

\bibitem[{\citenamefont{Baldsiefen and Gross}(2012)}]{Baldsiefen_al_1.2012}
\bibinfo{author}{\bibfnamefont{T.}~\bibnamefont{Baldsiefen}} \bibnamefont{and}
  \bibinfo{author}{\bibfnamefont{E.~K.~U.} \bibnamefont{Gross}}
  (\bibinfo{year}{2012}).

\bibitem[{\citenamefont{Baldsiefen et~al.}(2012)\citenamefont{Baldsiefen, Eich,
  and Gross}}]{Baldsiefen_al_2.2012}
\bibinfo{author}{\bibfnamefont{T.}~\bibnamefont{Baldsiefen}},
  \bibinfo{author}{\bibfnamefont{F.~G.} \bibnamefont{Eich}}, \bibnamefont{and}
  \bibinfo{author}{\bibfnamefont{E.~K.~U.} \bibnamefont{Gross}}
  (\bibinfo{year}{2012}).

\bibitem[{\citenamefont{Helbig et~al.}(2007)\citenamefont{Helbig, Lathiotakis,
  Albrecht, and Gross}}]{Helbig_al.2007}
\bibinfo{author}{\bibfnamefont{N.}~\bibnamefont{Helbig}},
  \bibinfo{author}{\bibfnamefont{N.~N.} \bibnamefont{Lathiotakis}},
  \bibinfo{author}{\bibfnamefont{M.}~\bibnamefont{Albrecht}}, \bibnamefont{and}
  \bibinfo{author}{\bibfnamefont{E.~K.~U.} \bibnamefont{Gross}},
  \bibinfo{journal}{Eur. Phys. Lett.} \textbf{\bibinfo{volume}{77}},
  \bibinfo{pages}{67003} (\bibinfo{year}{2007}),
  \urlprefix\url{http://dx.doi.org/10.1209/0295-5075/77/67003}.

\bibitem[{\citenamefont{Marques and
  Lathiotakis}(2008)}]{Marques_Lathiotakis.2008}
\bibinfo{author}{\bibfnamefont{M.~A.~L.} \bibnamefont{Marques}}
  \bibnamefont{and} \bibinfo{author}{\bibfnamefont{N.~N.}
  \bibnamefont{Lathiotakis}}, \bibinfo{journal}{Phys. Rev. A}
  \textbf{\bibinfo{volume}{77}}, \bibinfo{pages}{032509}
  (\bibinfo{year}{2008}),
  \urlprefix\url{http://dx.doi.org/10.1103/PhysRevA.77.032509}.

\bibitem[{\citenamefont{Piris et~al.}(2010)\citenamefont{Piris, Matxain, Lopez,
  and Ugalde}}]{Piris_al.2010}
\bibinfo{author}{\bibfnamefont{M.}~\bibnamefont{Piris}},
  \bibinfo{author}{\bibfnamefont{J.~M.} \bibnamefont{Matxain}},
  \bibinfo{author}{\bibfnamefont{X.}~\bibnamefont{Lopez}}, \bibnamefont{and}
  \bibinfo{author}{\bibfnamefont{J.~M.} \bibnamefont{Ugalde}},
  \bibinfo{journal}{J. Chem. Phys.} \textbf{\bibinfo{volume}{132}},
  \bibinfo{pages}{031103} (\bibinfo{year}{2010}),
  \urlprefix\url{http://dx.doi.org/10.1063/1.3298694}.

\bibitem[{\citenamefont{Helbig et~al.}(2009)\citenamefont{Helbig, Lathiotakis,
  and Gross}}]{Helbig_Lathiotakis_Gross.2009}
\bibinfo{author}{\bibfnamefont{N.}~\bibnamefont{Helbig}},
  \bibinfo{author}{\bibfnamefont{N.~N.} \bibnamefont{Lathiotakis}},
  \bibnamefont{and} \bibinfo{author}{\bibfnamefont{E.~K.~U.}
  \bibnamefont{Gross}}, \bibinfo{journal}{Phys. Rev. A}
  \textbf{\bibinfo{volume}{79}}, \bibinfo{pages}{022504}
  (\bibinfo{year}{2009}),
  \urlprefix\url{http://dx.doi.org/10.1103/PhysRevA.79.022504}.

\bibitem[{\citenamefont{Lathiotakis et~al.}(2010)\citenamefont{Lathiotakis,
  Sharma, Helbig, Dewhurst, Marques, Eich, Baldsiefen, Zacarias, and
  Gross}}]{Baldsiefen.2010}
\bibinfo{author}{\bibfnamefont{N.~N.} \bibnamefont{Lathiotakis}},
  \bibinfo{author}{\bibfnamefont{S.}~\bibnamefont{Sharma}},
  \bibinfo{author}{\bibfnamefont{N.}~\bibnamefont{Helbig}},
  \bibinfo{author}{\bibfnamefont{J.~K.} \bibnamefont{Dewhurst}},
  \bibinfo{author}{\bibfnamefont{M.~A.~L.} \bibnamefont{Marques}},
  \bibinfo{author}{\bibfnamefont{F.}~\bibnamefont{Eich}},
  \bibinfo{author}{\bibfnamefont{T.}~\bibnamefont{Baldsiefen}},
  \bibinfo{author}{\bibfnamefont{A.}~\bibnamefont{Zacarias}}, \bibnamefont{and}
  \bibinfo{author}{\bibfnamefont{E.~K.~U.} \bibnamefont{Gross}},
  \bibinfo{journal}{Z. Phys. Chem.} \textbf{\bibinfo{volume}{224}},
  \bibinfo{pages}{467} (\bibinfo{year}{2010}),
  \urlprefix\url{http://dx.doi.org/10.1524/zpch.2010.6118}.

\bibitem[{\citenamefont{Lathiotakis et~al.}(2009)\citenamefont{Lathiotakis,
  Sharma, Dewhurst, Eich, Marques, and Gross}}]{Lathiothakis_al.2009}
\bibinfo{author}{\bibfnamefont{N.~N.} \bibnamefont{Lathiotakis}},
  \bibinfo{author}{\bibfnamefont{S.}~\bibnamefont{Sharma}},
  \bibinfo{author}{\bibfnamefont{J.~K.} \bibnamefont{Dewhurst}},
  \bibinfo{author}{\bibfnamefont{F.~G.} \bibnamefont{Eich}},
  \bibinfo{author}{\bibfnamefont{M.~A.~L.} \bibnamefont{Marques}},
  \bibnamefont{and} \bibinfo{author}{\bibfnamefont{E.~K.~U.}
  \bibnamefont{Gross}}, \bibinfo{journal}{Phys. Rev. A}
  \textbf{\bibinfo{volume}{79}}, \bibinfo{pages}{040501}
  (\bibinfo{year}{2009}),
  \urlprefix\url{http://dx.doi.org/10.1103/PhysRevA.79.040501}.

\bibitem[{\citenamefont{Sharma et~al.}(2008)\citenamefont{Sharma, Dewhurst,
  Lathiotakis, and Gross}}]{Sharma_al.2008}
\bibinfo{author}{\bibfnamefont{S.}~\bibnamefont{Sharma}},
  \bibinfo{author}{\bibfnamefont{J.~K.} \bibnamefont{Dewhurst}},
  \bibinfo{author}{\bibfnamefont{N.~N.} \bibnamefont{Lathiotakis}},
  \bibnamefont{and} \bibinfo{author}{\bibfnamefont{E.~K.~U.}
  \bibnamefont{Gross}}, \bibinfo{journal}{Phys. Rev. B}
  \textbf{\bibinfo{volume}{78}}, \bibinfo{pages}{201103}
  (\bibinfo{year}{2008}),
  \urlprefix\url{http://dx.doi.org/10.1103/PhysRevB.78.201103}.

\bibitem[{\citenamefont{Singwi et~al.}(1968)\citenamefont{Singwi, Tosi, Land,
  and Sj\"{o}lander}}]{Singwi_Tosi_Land_Sjolander.1968}
\bibinfo{author}{\bibfnamefont{K.~S.} \bibnamefont{Singwi}},
  \bibinfo{author}{\bibfnamefont{M.~P.} \bibnamefont{Tosi}},
  \bibinfo{author}{\bibfnamefont{R.~H.} \bibnamefont{Land}}, \bibnamefont{and}
  \bibinfo{author}{\bibfnamefont{A.}~\bibnamefont{Sj\"{o}lander}},
  \bibinfo{journal}{Phys. Rev.} \textbf{\bibinfo{volume}{176}},
  \bibinfo{pages}{589} (\bibinfo{year}{1968}),
  \urlprefix\url{http://dx.doi.org/10.1103/PhysRev.176.589}.

\bibitem[{\citenamefont{Ceperley and Alder}(1980)}]{Ceperley_Alder.1980}
\bibinfo{author}{\bibfnamefont{D.~M.} \bibnamefont{Ceperley}} \bibnamefont{and}
  \bibinfo{author}{\bibfnamefont{B.~J.} \bibnamefont{Alder}},
  \bibinfo{journal}{Phys. Rev. Lett.} \textbf{\bibinfo{volume}{45}},
  \bibinfo{pages}{566} (\bibinfo{year}{1980}),
  \urlprefix\url{http://dx.doi.org/10.1103/PhysRevLett.45.566}.

\bibitem[{\citenamefont{Perdew and Wang}(1992)}]{Perdew_Wang.1992}
\bibinfo{author}{\bibfnamefont{J.~P.} \bibnamefont{Perdew}} \bibnamefont{and}
  \bibinfo{author}{\bibfnamefont{Y.}~\bibnamefont{Wang}},
  \bibinfo{journal}{Phys. Rev. B} \textbf{\bibinfo{volume}{45}},
  \bibinfo{pages}{13244} (\bibinfo{year}{1992}),
  \urlprefix\url{http://dx.doi.org/10.1103/PhysRevB.45.13244}.

\bibitem[{\citenamefont{Dharma-wardana and
  Perrot}(2000)}]{Dharma-wardana_Perrot.2000}
\bibinfo{author}{\bibfnamefont{M.~W.~C.} \bibnamefont{Dharma-wardana}}
  \bibnamefont{and} \bibinfo{author}{\bibfnamefont{F.}~\bibnamefont{Perrot}},
  \bibinfo{journal}{Phys. Rev. Lett.} \textbf{\bibinfo{volume}{84}},
  \bibinfo{pages}{959} (\bibinfo{year}{2000}),
  \urlprefix\url{http://dx.doi.org/10.1103/PhysRevLett.84.959}.

\bibitem[{\citenamefont{Tanaka and Ichimaru}(1986)}]{Tanaka_Ichimaru.1986}
\bibinfo{author}{\bibfnamefont{S.}~\bibnamefont{Tanaka}} \bibnamefont{and}
  \bibinfo{author}{\bibfnamefont{S.}~\bibnamefont{Ichimaru}},
  \bibinfo{journal}{J. Phys. Soc. Jpn.} \textbf{\bibinfo{volume}{55}},
  \bibinfo{pages}{2278} (\bibinfo{year}{1986}),
  \urlprefix\url{http://dx.doi.org/10.1143/JPSJ.55.2278}.

\bibitem[{\citenamefont{Schweng and B\"{o}hm}(1993)}]{Schweng_Boehm.1993}
\bibinfo{author}{\bibfnamefont{H.~K.} \bibnamefont{Schweng}} \bibnamefont{and}
  \bibinfo{author}{\bibfnamefont{H.~M.} \bibnamefont{B\"{o}hm}},
  \bibinfo{journal}{Phys. Rev. B} \textbf{\bibinfo{volume}{48}},
  \bibinfo{pages}{2037} (\bibinfo{year}{1993}),
  \urlprefix\url{http://dx.doi.org/10.1103/PhysRevB.48.2037}.

\bibitem[{\citenamefont{Iyetomi and Ichimaru}(1986)}]{Iyetomi_Ichimaru.1986}
\bibinfo{author}{\bibfnamefont{H.}~\bibnamefont{Iyetomi}} \bibnamefont{and}
  \bibinfo{author}{\bibfnamefont{S.}~\bibnamefont{Ichimaru}},
  \bibinfo{journal}{Phys. Rev. A} \textbf{\bibinfo{volume}{34}},
  \bibinfo{pages}{433} (\bibinfo{year}{1986}),
  \urlprefix\url{http://dx.doi.org/10.1103/PhysRevA.34.433}.

\bibitem[{\citenamefont{Tanaka and Ichimaru}(1989)}]{Tanaka_Ichimaru.1989}
\bibinfo{author}{\bibfnamefont{S.}~\bibnamefont{Tanaka}} \bibnamefont{and}
  \bibinfo{author}{\bibfnamefont{S.}~\bibnamefont{Ichimaru}},
  \bibinfo{journal}{Phys. Rev. B} \textbf{\bibinfo{volume}{39}},
  \bibinfo{pages}{1036} (\bibinfo{year}{1989}),
  \urlprefix\url{http://dx.doi.org/10.1103/PhysRevB.39.1036}.

\bibitem[{\citenamefont{Perrot and
  Dharma-wardana}(2000)}]{Perrot_Dharma-Wardana.2000}
\bibinfo{author}{\bibfnamefont{F.}~\bibnamefont{Perrot}} \bibnamefont{and}
  \bibinfo{author}{\bibfnamefont{M.~W.~C.} \bibnamefont{Dharma-wardana}},
  \bibinfo{journal}{Phys. Rev. B} \textbf{\bibinfo{volume}{62}},
  \bibinfo{pages}{16536} (\bibinfo{year}{2000}),
  \urlprefix\url{http://dx.doi.org/10.1103/PhysRevB.62.16536}.

\bibitem[{\citenamefont{Zong et~al.}(2002)\citenamefont{Zong, Lin, and
  Ceperley}}]{Zong_Lin_Ceperley.2002}
\bibinfo{author}{\bibfnamefont{F.~H.} \bibnamefont{Zong}},
  \bibinfo{author}{\bibfnamefont{C.}~\bibnamefont{Lin}}, \bibnamefont{and}
  \bibinfo{author}{\bibfnamefont{D.~M.} \bibnamefont{Ceperley}},
  \bibinfo{journal}{Phys. Rev. E} \textbf{\bibinfo{volume}{66}},
  \bibinfo{pages}{036703} (\bibinfo{year}{2002}),
  \urlprefix\url{http://dx.doi.org/10.1103/PhysRevE.66.036703}.

\bibitem[{\citenamefont{Conduit et~al.}(2009)\citenamefont{Conduit, Green, and
  Simons}}]{Conduit_Green_Simons.2009}
\bibinfo{author}{\bibfnamefont{G.~J.} \bibnamefont{Conduit}},
  \bibinfo{author}{\bibfnamefont{A.~G.} \bibnamefont{Green}}, \bibnamefont{and}
  \bibinfo{author}{\bibfnamefont{B.~D.} \bibnamefont{Simons}},
  \bibinfo{journal}{Phys. Rev. Lett.} \textbf{\bibinfo{volume}{103}},
  \bibinfo{pages}{207201} (\bibinfo{year}{2009}),
  \urlprefix\url{http://dx.doi.org/10.1103/PhysRevLett.103.207201}.

\bibitem[{\citenamefont{Perrot}(1979)}]{Perrot.1979}
\bibinfo{author}{\bibfnamefont{F.}~\bibnamefont{Perrot}},
  \bibinfo{journal}{Phys. Rev. A} \textbf{\bibinfo{volume}{20}},
  \bibinfo{pages}{586} (\bibinfo{year}{1979}),
  \urlprefix\url{http://dx.doi.org/10.1103/PhysRevA.20.586}.

\bibitem[{\citenamefont{Gupta and
  Rajagopal}(1980{\natexlab{a}})}]{Gupta_Rajagopal_2.1980}
\bibinfo{author}{\bibfnamefont{U.}~\bibnamefont{Gupta}} \bibnamefont{and}
  \bibinfo{author}{\bibfnamefont{A.~K.} \bibnamefont{Rajagopal}},
  \bibinfo{journal}{Phys. Rev. A} \textbf{\bibinfo{volume}{22}},
  \bibinfo{pages}{2792} (\bibinfo{year}{1980}{\natexlab{a}}),
  \urlprefix\url{http://dx.doi.org/10.1103/PhysRevA.22.2792}.

\bibitem[{\citenamefont{Dharma-wardana and
  Taylor}(1981)}]{Dharma-Wardana_Taylor.1981}
\bibinfo{author}{\bibfnamefont{M.~W.~C.} \bibnamefont{Dharma-wardana}}
  \bibnamefont{and} \bibinfo{author}{\bibfnamefont{R.}~\bibnamefont{Taylor}},
  \bibinfo{journal}{J. Phys. C} \textbf{\bibinfo{volume}{14}},
  \bibinfo{pages}{629} (\bibinfo{year}{1981}),
  \urlprefix\url{http://dx.doi.org/10.1088/0022-3719/14/5/011}.

\bibitem[{\citenamefont{Gupta and Rajagopal}(1982)}]{Gupta_Rajagopal.1982}
\bibinfo{author}{\bibfnamefont{U.}~\bibnamefont{Gupta}} \bibnamefont{and}
  \bibinfo{author}{\bibfnamefont{A.~K.} \bibnamefont{Rajagopal}},
  \bibinfo{journal}{Phys. Rep.} \textbf{\bibinfo{volume}{87}},
  \bibinfo{pages}{259} (\bibinfo{year}{1982}),
  \urlprefix\url{http://dx.doi.org/10.1016/0370-1573(82)90077-1}.

\bibitem[{\citenamefont{Perrot and
  Dharma-wardana}(1984)}]{Perrot_Dharma-Wardana.1984}
\bibinfo{author}{\bibfnamefont{F.}~\bibnamefont{Perrot}} \bibnamefont{and}
  \bibinfo{author}{\bibfnamefont{M.~W.~C.} \bibnamefont{Dharma-wardana}},
  \bibinfo{journal}{Phys. Rev. A} \textbf{\bibinfo{volume}{30}},
  \bibinfo{pages}{2619} (\bibinfo{year}{1984}),
  \urlprefix\url{http://dx.doi.org/10.1103/PhysRevA.30.2619}.

\bibitem[{\citenamefont{Kanhere et~al.}(1986)\citenamefont{Kanhere, Panat,
  Rajagopal, and Callaway}}]{Kanhere_Panat_Rajagopal_Callaway.1986}
\bibinfo{author}{\bibfnamefont{D.~G.} \bibnamefont{Kanhere}},
  \bibinfo{author}{\bibfnamefont{P.~V.} \bibnamefont{Panat}},
  \bibinfo{author}{\bibfnamefont{A.~K.} \bibnamefont{Rajagopal}},
  \bibnamefont{and} \bibinfo{author}{\bibfnamefont{J.}~\bibnamefont{Callaway}},
  \bibinfo{journal}{Phys. Rev. A} \textbf{\bibinfo{volume}{33}},
  \bibinfo{pages}{490} (\bibinfo{year}{1986}),
  \urlprefix\url{http://dx.doi.org/10.1103/PhysRevA.33.490}.

\bibitem[{\citenamefont{Dandrea et~al.}(1986)\citenamefont{Dandrea, Ashcroft,
  and Carlsson}}]{Dandrea_Ashcroft_Carlsson.1986}
\bibinfo{author}{\bibfnamefont{R.~G.} \bibnamefont{Dandrea}},
  \bibinfo{author}{\bibfnamefont{N.~W.} \bibnamefont{Ashcroft}},
  \bibnamefont{and} \bibinfo{author}{\bibfnamefont{A.~E.}
  \bibnamefont{Carlsson}}, \bibinfo{journal}{Phys. Rev. B}
  \textbf{\bibinfo{volume}{34}}, \bibinfo{pages}{2097} (\bibinfo{year}{1986}),
  \urlprefix\url{http://dx.doi.org/10.1103/PhysRevB.34.2097}.

\bibitem[{\citenamefont{L\"{o}wdin}(1955)}]{Loewdin.1955}
\bibinfo{author}{\bibfnamefont{P.~O.} \bibnamefont{L\"{o}wdin}},
  \bibinfo{journal}{Phys. Rev.} \textbf{\bibinfo{volume}{97}},
  \bibinfo{pages}{1474} (\bibinfo{year}{1955}),
  \urlprefix\url{http://dx.doi.org/10.1103/PhysRev.97.1474}.

\bibitem[{\citenamefont{Kurth et~al.}(1999)\citenamefont{Kurth, Marques,
  L\"{u}ders, and Gross}}]{Kurth_Marques_Lueders_Gross.1999}
\bibinfo{author}{\bibfnamefont{S.}~\bibnamefont{Kurth}},
  \bibinfo{author}{\bibfnamefont{M.}~\bibnamefont{Marques}},
  \bibinfo{author}{\bibfnamefont{M.}~\bibnamefont{L\"{u}ders}},
  \bibnamefont{and} \bibinfo{author}{\bibfnamefont{E.~K.~U.}
  \bibnamefont{Gross}}, \bibinfo{journal}{Phys. Rev. Lett.}
  \textbf{\bibinfo{volume}{83}}, \bibinfo{pages}{2628} (\bibinfo{year}{1999}),
  \urlprefix\url{http://dx.doi.org/10.1103/PhysRevLett.83.2628}.

\bibitem[{\citenamefont{Ullrich and Gross}(1996)}]{Ullrich_Gross.1996}
\bibinfo{author}{\bibfnamefont{C.~A.} \bibnamefont{Ullrich}} \bibnamefont{and}
  \bibinfo{author}{\bibfnamefont{E.~K.~U.} \bibnamefont{Gross}},
  \bibinfo{journal}{Aust. J. Phys.} \textbf{\bibinfo{volume}{49}},
  \bibinfo{pages}{103} (\bibinfo{year}{1996}).

\bibitem[{\citenamefont{Ceperley}(1995)}]{Ceperley.1995}
\bibinfo{author}{\bibfnamefont{D.~M.} \bibnamefont{Ceperley}},
  \bibinfo{journal}{Rev. Mod. Phys.} \textbf{\bibinfo{volume}{67}},
  \bibinfo{pages}{279} (\bibinfo{year}{1995}),
  \urlprefix\url{http://dx.doi.org/10.1103/RevModPhys.67.279}.

\bibitem[{\citenamefont{Troyer and Wiese}(2005)}]{Troyer_Wiese.2005}
\bibinfo{author}{\bibfnamefont{M.}~\bibnamefont{Troyer}} \bibnamefont{and}
  \bibinfo{author}{\bibfnamefont{U.~J.} \bibnamefont{Wiese}},
  \bibinfo{journal}{Phys. Rev. Lett.} \textbf{\bibinfo{volume}{94}},
  \bibinfo{pages}{170201} (\bibinfo{year}{2005}),
  \urlprefix\url{http://dx.doi.org/10.1103/PhysRevLett.94.170201}.

\bibitem[{\citenamefont{M\"{u}ller}(1984)}]{Mueller.1984}
\bibinfo{author}{\bibfnamefont{A.}~\bibnamefont{M\"{u}ller}},
  \bibinfo{journal}{Phys. Lett. A} \textbf{\bibinfo{volume}{105}},
  \bibinfo{pages}{446} (\bibinfo{year}{1984}),
  \urlprefix\url{http://dx.doi.org/10.1016/0375-9601(84)91034-X}.

\bibitem[{\citenamefont{Buijse and Baerends}(2002)}]{Buijse_Baerends.2002}
\bibinfo{author}{\bibnamefont{Buijse}} \bibnamefont{and}
  \bibinfo{author}{\bibnamefont{Baerends}}, \bibinfo{journal}{Mol. Phys.}
  \textbf{\bibinfo{volume}{100}}, \bibinfo{pages}{401} (\bibinfo{year}{2002}),
  \urlprefix\url{http://dx.doi.org/10.1080/00268970110070243}.

\bibitem[{\citenamefont{Gritsenko et~al.}(2005)\citenamefont{Gritsenko, Pernal,
  and Baerends}}]{Gritsenko_Pernal_Baerends.2005}
\bibinfo{author}{\bibfnamefont{O.}~\bibnamefont{Gritsenko}},
  \bibinfo{author}{\bibfnamefont{K.}~\bibnamefont{Pernal}}, \bibnamefont{and}
  \bibinfo{author}{\bibfnamefont{E.~J.} \bibnamefont{Baerends}},
  \bibinfo{journal}{J. Chem. Phys.} \textbf{\bibinfo{volume}{122}},
  \bibinfo{pages}{204102} (\bibinfo{year}{2005}),
  \urlprefix\url{http://dx.doi.org/10.1063/1.1906203}.

\bibitem[{\citenamefont{Lathiotakis and
  Marques}(2008)}]{Lathiotakis_Marques.2008}
\bibinfo{author}{\bibfnamefont{N.~N.} \bibnamefont{Lathiotakis}}
  \bibnamefont{and} \bibinfo{author}{\bibfnamefont{M.~A.~L.}
  \bibnamefont{Marques}}, \bibinfo{journal}{J. Chem. Phys.}
  \textbf{\bibinfo{volume}{128}}, \bibinfo{pages}{184103+}
  (\bibinfo{year}{2008}), \urlprefix\url{http://dx.doi.org/10.1063/1.2899328}.

\bibitem[{\citenamefont{Ayers et~al.}(1998)\citenamefont{Ayers, Day, and
  Morrison}}]{Ayers_al.1998}
\bibinfo{author}{\bibfnamefont{P.~W.} \bibnamefont{Ayers}},
  \bibinfo{author}{\bibfnamefont{O.~W.} \bibnamefont{Day}}, \bibnamefont{and}
  \bibinfo{author}{\bibfnamefont{R.~C.} \bibnamefont{Morrison}},
  \bibinfo{journal}{Int. J. Quantum Chem.} \textbf{\bibinfo{volume}{69}},
  \bibinfo{pages}{541} (\bibinfo{year}{1998}),
  \urlprefix\url{http://dx.doi.org/10.1002/(SICI)1097-461X(1998)69:4\%3C541::AID-QUA11\%3E3.0.CO;2-2}.

\bibitem[{\citenamefont{Gori-Giorgi and Ziesche}(2002)}]{Giorgi_Ziesche.2002}
\bibinfo{author}{\bibfnamefont{P.}~\bibnamefont{Gori-Giorgi}} \bibnamefont{and}
  \bibinfo{author}{\bibfnamefont{P.}~\bibnamefont{Ziesche}},
  \bibinfo{journal}{Phys. Rev. B} \textbf{\bibinfo{volume}{66}},
  \bibinfo{pages}{235116} (\bibinfo{year}{2002}),
  \urlprefix\url{http://dx.doi.org/10.1103/PhysRevB.66.235116}.

\bibitem[{\citenamefont{Kimball}(1975)}]{Kimball.1975}
\bibinfo{author}{\bibfnamefont{J.~C.} \bibnamefont{Kimball}},
  \bibinfo{journal}{J. Phys. A: Math. Gen.} \textbf{\bibinfo{volume}{8}},
  \bibinfo{pages}{1513} (\bibinfo{year}{1975}),
  \urlprefix\url{http://dx.doi.org/10.1088/0305-4470/8/9/021}.

\bibitem[{\citenamefont{Friesecke}(2003)}]{Friesecke.2003}
\bibinfo{author}{\bibfnamefont{G.}~\bibnamefont{Friesecke}},
  \bibinfo{journal}{Proc. R. Soc. London A} \textbf{\bibinfo{volume}{459}},
  \bibinfo{pages}{47} (\bibinfo{year}{2003}),
  \urlprefix\url{http://dx.doi.org/10.1098/rspa.2002.1027}.

\bibitem[{\citenamefont{Lathiotakis et~al.}(2007)\citenamefont{Lathiotakis,
  Helbig, and Gross}}]{Lathiotakis_Helbig_Gross.2007}
\bibinfo{author}{\bibfnamefont{N.~N.} \bibnamefont{Lathiotakis}},
  \bibinfo{author}{\bibfnamefont{N.}~\bibnamefont{Helbig}}, \bibnamefont{and}
  \bibinfo{author}{\bibfnamefont{E.~K.~U.} \bibnamefont{Gross}},
  \bibinfo{journal}{Phys. Rev. B} \textbf{\bibinfo{volume}{75}},
  \bibinfo{pages}{195120} (\bibinfo{year}{2007}),
  \urlprefix\url{http://dx.doi.org/10.1103/PhysRevB.75.195120}.

\bibitem[{\citenamefont{Attaccalite et~al.}(2002)\citenamefont{Attaccalite,
  Moroni, Gori-Giorgi, and Bachelet}}]{Attaccalite_al.2002}
\bibinfo{author}{\bibfnamefont{C.}~\bibnamefont{Attaccalite}},
  \bibinfo{author}{\bibfnamefont{S.}~\bibnamefont{Moroni}},
  \bibinfo{author}{\bibfnamefont{P.}~\bibnamefont{Gori-Giorgi}},
  \bibnamefont{and} \bibinfo{author}{\bibfnamefont{G.~B.}
  \bibnamefont{Bachelet}}, \bibinfo{journal}{Phys. Rev. Lett.}
  \textbf{\bibinfo{volume}{88}}, \bibinfo{pages}{256601}
  (\bibinfo{year}{2002}),
  \urlprefix\url{http://dx.doi.org/10.1103/PhysRevLett.88.256601}.

\bibitem[{\citenamefont{Gupta and
  Rajagopal}(1980{\natexlab{b}})}]{Gupta_Rajagopal_1.1980}
\bibinfo{author}{\bibfnamefont{U.}~\bibnamefont{Gupta}} \bibnamefont{and}
  \bibinfo{author}{\bibfnamefont{A.~K.} \bibnamefont{Rajagopal}},
  \bibinfo{journal}{Phys. Rev. A} \textbf{\bibinfo{volume}{21}},
  \bibinfo{pages}{2064} (\bibinfo{year}{1980}{\natexlab{b}}),
  \urlprefix\url{http://dx.doi.org/10.1103/PhysRevA.21.2064}.

\bibitem[{\citenamefont{Ismail-Beigi}(2010)}]{Ismail-Beigi.2010}
\bibinfo{author}{\bibfnamefont{S.}~\bibnamefont{Ismail-Beigi}},
  \bibinfo{journal}{Phys. Rev. B} \textbf{\bibinfo{volume}{81}},
  \bibinfo{pages}{195126} (\bibinfo{year}{2010}),
  \urlprefix\url{http://dx.doi.org/10.1103/PhysRevB.81.195126}.

\bibitem[{\citenamefont{Hedin}(1965)}]{Hedin.1965}
\bibinfo{author}{\bibfnamefont{L.}~\bibnamefont{Hedin}},
  \bibinfo{journal}{Phys. Rev.} \textbf{\bibinfo{volume}{139}}
  (\bibinfo{year}{1965}),
  \urlprefix\url{http://dx.doi.org/10.1103/PhysRev.139.A796}.

\end{thebibliography}
\end{document}